\documentclass[aps,showpacs]{revtex4}

\usepackage{amssymb,amsmath}
\usepackage[dvips]{graphicx}

\newcommand{\bd}[1]{\mbox{\boldmath$#1$}}

\begin{document}

\title{\mbox{Phase synchronization between collective rhythms of globally coupled oscillator groups:} noisy identical case}

\author{Yoji Kawamura}
\email{ykawamura@jamstec.go.jp}
\affiliation{Institute for Research on Earth Evolution,
Japan Agency for Marine-Earth Science and Technology, Yokohama 236-0001, Japan}

\author{Hiroya Nakao}
\affiliation{Department of Physics, Kyoto University, Kyoto 606-8502, Japan}
\affiliation{JST, CREST, Kyoto 606-8502, Japan}

\author{Kensuke Arai}
\affiliation{Brain Science Institute, RIKEN, Wako 351-0198, Japan}

\author{Hiroshi Kori}
\affiliation{Division of Advanced Sciences, Ochadai Academic Production, Ochanomizu University, Tokyo 112-8610, Japan}
\affiliation{PRESTO, Japan Science and Technology Agency, Kawaguchi 332-0012, Japan}

\author{Yoshiki Kuramoto}
\affiliation{Research Institute for Mathematical Sciences, Kyoto University, Kyoto 606-8502, Japan}
\affiliation{Institute for Integrated Cell-Material Sciences, Kyoto University, Kyoto 606-8501, Japan}

\date{July 26, 2010}


\begin{abstract}
  We theoretically investigate collective phase synchronization between interacting groups
  of globally coupled noisy identical phase oscillators exhibiting macroscopic rhythms.
  Using the phase reduction method, we derive coupled collective phase equations
  describing the macroscopic rhythms of the groups from microscopic Langevin phase equations of the individual oscillators
  via nonlinear Fokker-Planck equations.
  For sinusoidal microscopic coupling,
  we determine the type of the collective phase coupling function,
  i.e., whether the groups exhibit in-phase or anti-phase synchronization.
  We show that the macroscopic rhythms can exhibit effective anti-phase synchronization
  even if the microscopic phase coupling between the groups is in-phase, and vice versa.
  Moreover, near the onset of collective oscillations,
  we analytically obtain the collective phase coupling function
  using center-manifold and phase reductions of the nonlinear Fokker-Planck equations.
\end{abstract}

\maketitle

\begin{quotation}
{\bf
Systems of limit-cycle oscillators are used to model various rhythmic phenomena
in natural and artificial systems.
When the interaction between the oscillators is weak,
the system can generally be described as coupled phase oscillators.
Qualitative understanding of synchronization phenomena,
in particular the emergence of collective oscillations,
has been achieved successfully through this type of reduced model.
In the present paper, we develop collective phase reduction method for macroscopic oscillations
arising from mutual synchronization of many microscopic oscillators within a group,
which is typical in the real world.
We show that weakly interacting groups of oscillators,
each consisting of globally coupled noisy identical phase oscillators
and exhibiting collective oscillations,
can be reduced to coupled equations for collective phases.
This makes it possible to analyze the nature of synchronization between the collective oscillations
in a closed way at the macroscopic level.
}
\end{quotation}

\section{Introduction} \label{sec:introduction}


Populations of coupled rhythmic elements can exhibit macroscopic oscillations through mutual synchronization
~\cite{ref:winfree80,ref:kuramoto84,ref:pikovsky01,ref:strogatz03,ref:manrubia04,ref:izhikevich07}.
The phase oscillator models have played important roles in theoretically analyzing their behavior,
and the special class of models given by globally coupled phase oscillators in particular,
was studied most intensively in the past
~\cite{ref:strogatz00,ref:acebron05,ref:boccaletti06,ref:sagues07,ref:arenas08,ref:dorogovtsev08}.
Theoretical predictions based on such models have also been experimentally validated,
e.g., in electrochemical oscillator systems~\cite{ref:kiss02,ref:kiss05,ref:kiss07,ref:kiss08,ref:kori08}
and in discrete chemical oscillator populations~\cite{ref:toth06,ref:taylor09,ref:tinsley10}.

Recently, macroscopic synchronization between interacting groups of globally coupled phase oscillators
exhibiting collective oscillations has attracted attention
~\cite{ref:okuda91,ref:montbrio04,ref:barreto08,ref:ott08,ref:sheeba08,ref:sheeba09}.
In most of the works so far, the macroscopic properties such as mutual entrainment between the groups
have been analyzed through the microscopic individual phases.
However, because we are interested in the macroscopic behavior of the collective rhythms exhibited by the oscillator groups,
it should be much more convenient if each group of oscillators can be treated as a single macroscopic oscillator.
Based on such consideration, we have developed collective phase reduction methods~\cite{ref:kawamura07,ref:kawamura08,ref:kori09},
which provide us with the collective phase sensitivity of macroscopic rhythms of the oscillator group to weak perturbations.

In this paper, we employ the notion of collective phase description~\cite{ref:kawamura07,ref:kawamura08,ref:kori09},
and formulate a theory for weakly interacting groups of globally coupled noisy identical phase oscillators
in a closed form at the macroscopic level.
Specifically, we derive coupled collective phase equations from microscopic Langevin phase equations
describing weakly interacting groups of globally coupled phase oscillators.
A general formula that gives collective phase coupling functions
from the microscopic phase coupling functions between the individual oscillators is obtained,
and for the case with sinusoidal coupling, the types of the collective phase coupling function
are determined as a function of the coupling parameters.
Near the onset of collective oscillations,
we can even analytically obtain the collective phase coupling function by the center-manifold and phase reductions.
Based on the collective phase equations, we illustrate counter-intuitive phenomena in which
two oscillator groups become anti-phase synchronized in spite of in-phase microscopic coupling between the groups,
and vice versa (in-phase synchronization despite anti-phase microscopic coupling).

In Ref.~\cite{ref:kawamura10}, we considered a similar problem, namely,
collective phase synchronization between two groups of globally coupled oscillators.
The crucial difference is that we treat noisy identical phase oscillators in the present work,
whereas we analyzed noiseless non-identical phase oscillators in Ref.~\cite{ref:kawamura10}.
Although these two situations look similar,
they are essentially different physical systems (i.e., stochastic vs. deterministic)
and mathematical treatments should be developed independently.
Here we apply center-manifold reduction as well as phase reduction
to nonlinear Fokker-Planck equations governing the oscillator groups,
whereas we used the Ott-Antonsen ansatz in the analysis of the noiseless non-identical system~\cite{ref:kawamura10}.
In both cases, despite the large difference in their mathematical structures,
we obtain similar coupled collective phase equations describing macroscopic dynamics of the groups.
Thus, the present paper and Ref.~\cite{ref:kawamura10} are mutually complementary
and together give deeper understanding of macroscopic collective phenomena.

The organization of the present paper is the following.
In Sec.~\ref{sec:model},
we introduce a model of weakly interacting groups of globally coupled noisy phase oscillators
and illustrate both effective anti-phase and in-phase collective phase synchronization
between the groups by numerical simulations.
In Sec.~\ref{sec:phase},
we develop a theory that derives coupled collective phase equations from the microscopic model
and determine the effective type of phase coupling between collective oscillations.
In Sec.~\ref{sec:amplitude},
we analytically obtain the collective phase coupling function near the onset of collective oscillations
and discuss several important cases.
In Sec.~\ref{sec:turbulence},
we discuss a relation to noise-induced turbulence in a system of nonlocally coupled oscillators.
Concluding remarks will be given in the final section.

\section{The model and its dynamics} \label{sec:model}

\subsection{Interacting groups of globally coupled phase oscillators}

We consider two interacting groups of globally coupled noisy identical phase oscillators
described by the following model:
\begin{equation}
  \dot{\phi}_j^{(\sigma)}\left( t \right) = \omega + \frac{1}{N}
  \sum_{k=1}^N \Gamma\left( \phi_j^{(\sigma)} - \phi_k^{(\sigma)} \right)
  + \sqrt{D} \, \xi_j^{(\sigma)}\left( t \right) + \frac{\epsilon}{N}
  \sum_{k=1}^N \Gamma_{\sigma\tau}\left( \phi_j^{(\sigma)} - \phi_k^{(\tau)} \right)
  \label{eq:model}
\end{equation}
for $j = 1, \cdots, N$ and $(\sigma, \tau) = (1, 2)$ or $(2, 1)$,
where $\phi_j^{(\sigma)}(t)$ is the phase of the $j$-th oscillator in the $\sigma$-th group consisting of $N$ oscillators
and $\omega$ is the natural frequency common to all oscillators.
The second term on the right-hand side represents the internal coupling between the oscillators within the same group,
the third term represents the noise,
and the last term gives the external coupling between the oscillators belonging to different groups.
The internal phase coupling function $\Gamma(\phi)$ is assumed to be in-phase,
$d\Gamma(\phi)/d\phi|_{\phi=0} < 0$~\cite{ref:kuramoto84}, namely,
the oscillators within the same group tend to synchronize with each other.
The external phase coupling function between the groups is described by $\Gamma_{\sigma\tau}(\phi)$.
Characteristic intensity of the internal coupling within each group is scaled to unity,
whereas that of the external coupling between the groups is given by $\epsilon \geq 0$.
The noise $\xi_j^{(\sigma)}(t)$ is assumed to be white Gaussian~\cite{ref:horsthemke84,ref:risken89,ref:gardiner97},
whose statistics are given by
\begin{equation}
  \left\langle \xi_j^{(\sigma)}(t) \right\rangle
  = 0, \qquad
  \left\langle \xi_j^{(\sigma)}(t) \xi_k^{(\tau)}(s) \right\rangle
  = 2 \delta_{jk} \delta_{\sigma\tau} \delta(t - s).
\end{equation}
The noise intensity is characterized by $D \geq 0$.
When the external coupling is absent, i.e., $\epsilon = 0$,
Eq.~(\ref{eq:model}) has a critical noise intensity $D_{\rm c}$ below which phase coherent states are realized,
namely, collective oscillations arise when $0 \leq D < D_{\rm c}$~\cite{ref:kuramoto84}.
In the following, we assume that $\epsilon$ is sufficiently small
and each group of oscillators exhibits stable collective oscillations.

\subsection{Phase synchronization between collective oscillations}

In the following numerical simulations, we assume that the phase coupling functions are sinusoidal
(note however that our theory itself can be applied to general $2\pi$-periodic phase coupling functions~\cite{ref:kawamura08}).
Without loss of generality, the natural frequency can be assumed to be zero, $\omega = 0$.
The internal phase coupling function between the oscillators within the same group is given by
\begin{equation}
  \Gamma\left( \phi \right) = - \sin\left( \phi + \alpha \right), \qquad
  \left| \alpha \right| < \frac{\pi}{2},
  \label{eq:alpha}
\end{equation}
which is in-phase (attractive).
In this case, the critical noise intensity is given by $D_{\rm c} = (\cos \alpha)/2$ as explained in Sec.~\ref{sec:amplitude}.
The external phase coupling function is described by
\begin{equation}
  \Gamma_{\sigma\tau}\left( \phi \right) = - \sin\left( \phi + \beta \right), \qquad
  \left| \beta \right| \leq \pi,
  \label{eq:beta}
\end{equation}
which can be either in-phase (attractive) ($|\beta| < \pi/2$) or anti-phase (repulsive) ($|\beta| > \pi/2$)~\cite{ref:kuramoto84}.
Introducing a complex order parameter $A^{(\sigma)}(t)$
with modulus $R^{(\sigma)}(t)$ and phase $\Theta^{(\sigma)}(t)$ through
\begin{equation}
  A^{(\sigma)}(t) = R^{(\sigma)}(t) e^{i \Theta^{(\sigma)}(t)}
  = \frac{1}{N} \sum_{k=1}^N e^{i \phi^{(\sigma)}_k(t)},
  \label{eq:order}
\end{equation}
we can rewrite Eq.~(\ref{eq:model}) with the sinusoidal coupling functions
given in Eqs.~(\ref{eq:alpha}) and (\ref{eq:beta}) as follows:
\begin{equation}
  \dot{\phi}_j^{(\sigma)}(t) = \omega
  - R^{(\sigma)} \sin\left( \phi_j^{(\sigma)} - \Theta^{(\sigma)} + \alpha \right)
  + \sqrt{D} \, \xi_j^{(\sigma)}(t)
  - \epsilon R^{(\tau)} \sin\left( \phi_j^{(\sigma)} - \Theta^{(\tau)} + \beta \right).
  \label{eq:simulation}
\end{equation}
Note that $R^{(\sigma)}$ quantifies the degree of synchronization
and $\Theta^{(\sigma)}$ gives the {\it collective phase} of the $\sigma$-th group.

Focusing on weakly coupled collective oscillations,
we carried out numerical simulations of Eq.~(\ref{eq:simulation}) with Eq.~(\ref{eq:order}) under the following conditions:
the external coupling was assumed to be much weaker than the internal coupling, i.e., $\epsilon = 0.01 \ll 1$;
we set $D = D_{\rm c} / 2 = (\cos \alpha) / 4$ and $\alpha = 3 \pi / 8$;
the number of oscillators in each group was $N = 10000$, which was sufficiently large to observe clear collective oscillations.
We separately prepared two groups of phase oscillators exhibiting collective oscillations
and used these states as the initial conditions of the simulations.

In Fig.~\ref{fig:1}(a), evolution of the collective phase difference $| \Theta^{(1)} - \Theta^{(2)} |$
from almost in-phase synchronized state of the groups is shown.
In spite of the in-phase external phase coupling condition between individual oscillator pairs, $\beta = 3 \pi / 8$,
the collective phase difference $| \Theta^{(1)} - \Theta^{(2)} |$ eventually approached $\pi$, namely,
the two groups became anti-phase synchronized after some time.
Thus, Fig.~\ref{fig:1}(a) indicates that the collective phase coupling function between the group is anti-phase
although microscopic external phase coupling functions are in-phase.
In contrast, Fig.~\ref{fig:1}(b) shows evolution of $| \Theta^{(1)} - \Theta^{(2)} |$
from almost anti-phase synchronized state of the groups
with anti-phase microscopic external phase coupling function, $\beta = -5 \pi / 8$,
which eventually became in-phase synchronized.

Snapshots of the microscopic phase variables
after the collective phase difference has reached the asymptotic value in Fig.~\ref{fig:1}
are displayed in Fig.~\ref{fig:2}.
In Fig.~\ref{fig:2}(a), the two distributions of the oscillators are shifted by $\pi$,
indicating anti-phase synchronization between the groups.
In contrast, the two distributions almost overlap in Fig.~\ref{fig:2}(b),
i.e., they are in-phase synchronized.
Note that oscillators from different groups do not synchronize with each other,
in other words, the collective phase synchronization between the groups is not
due to complete synchronization of individual oscillators at the microscopic level.

Thus, the type of the collective phase coupling functions can be effectively different from
that of the microscopic external phase coupling functions,
depending on the collective dynamics of the oscillators taking place in each group.
We develop a theory that yields the collective phase coupling function
from the microscopic model in the following sections.

\section{Collective phase reduction} \label{sec:phase}

We derive coupled dynamical equations for the collective phase variables of the groups
from the Langevin phase equations of individual oscillators through nonlinear Fokker-Planck equations
and obtain a formula that relates the collective phase coupling function between the groups
to the microscopic phase coupling function between individual oscillator pairs from different groups.
Using them, we determine the type of the collective phase coupling function
and explain the results of the numerical simulations in Sec.~\ref{sec:model}.

\subsection{Nonlinear Fokker-Planck equations}

In the continuum limit, i.e., $N \to \infty$, the Langevin phase equations~(\ref{eq:model}) can be transformed into
the following coupled nonlinear Fokker-Planck equations~\cite{ref:kuramoto84,ref:okuda91,ref:kawamura07,ref:kawamura08}:
\begin{align}
  \frac{\partial}{\partial t} f^{(\sigma)}\left( \phi, t \right)
  = &- \frac{\partial}{\partial \phi} \left[ \left\{
    \omega + \int_{0}^{2\pi} d\phi' \, \Gamma\left( \phi - \phi' \right)
    f^{(\sigma)}\left( \phi', t \right) \right\}
    f^{(\sigma)}\left( \phi, t \right) \right]
  + D \frac{\partial^2}{\partial \phi^2} f^{(\sigma)}\left( \phi, t \right) \nonumber \\
  &- \epsilon \frac{\partial}{\partial \phi}
  \left[ \int_{0}^{2\pi} d\phi' \, \Gamma_{\sigma\tau}\left( \phi - \phi' \right)
    f^{(\tau)}\left( \phi', t \right) f^{(\sigma)}\left( \phi, t \right) \right],
  \label{eq:nfp}
\end{align}
for $(\sigma, \tau) = (1, 2)$ or $(2, 1)$.
Here, $f^{(\sigma)}(\phi, t)$ is the one-body probability density function
of the individual oscillator phase $\phi$ in the $\sigma$-th group,
which is normalized as $\int_{0}^{2\pi} d\phi f^{(\sigma)}\left( \phi, t \right) = 1$.
The first two terms on the right-hand side represent internal dynamics of the $\sigma$-th group,
and the third term represents weak interaction between $\sigma$-th group and $\tau$-th group.
The complex order parameter of Eq.~(\ref{eq:order}) is now expressed as
\begin{equation}
  A^{(\sigma)}(t) = R^{(\sigma)}(t) e^{i \Theta^{(\sigma)}(t)}
  = \int_{0}^{2\pi} d\phi \, e^{i \phi} f^{(\sigma)}\left( \phi, t \right).
\end{equation}
When the external coupling between the groups is absent,
each group of oscillators obeying Eq.~(\ref{eq:nfp}) with $\epsilon = 0$
exhibits collective rhythms under the condition
$0 \leq D < D_{\rm c}$~\cite{ref:kuramoto84,ref:okuda91,ref:kawamura07,ref:kawamura08}.
We assume that this situation persists even if $\epsilon$ becomes slightly positive
and the two groups interact with each other weakly.

The collectively oscillating solution of the nonlinear Fokker-Planck equations~(\ref{eq:nfp})
without external coupling ($\epsilon = 0$) can be expressed as a steadily rotating wave packet on a periodic interval $[0, 2\pi]$,
\begin{equation}
  f^{(\sigma)}\left( \phi, t \right) = f_0\left( \varphi^{(\sigma)} \right), \qquad
  \varphi^{(\sigma)} = \phi - \Theta^{(\sigma)}, \qquad
  \dot{\Theta}^{(\sigma)} = \Omega,
  \label{eq:f0}
\end{equation}
for $\sigma = 1, 2$, where the $f_0(\varphi)$ represents the steady functional shape of the wave packet,
$\Theta^{(\sigma)}$ is the location of the wave packet at time $t$, namely, the collective phase of the $\sigma$-th group,
and $\Omega$ is the collective frequency common to both groups~\cite{ref:kawamura07,ref:kawamura08}.

\subsection{Collective phase equations}

Let us assume $\epsilon = 0$ and focus on a single group.
The group index $\sigma$ will be dropped for the moment.
Inserting Eq.~(\ref{eq:f0}) into the nonlinear Fokker-Planck equation~(\ref{eq:nfp}) with $\epsilon = 0$,
we find that $f_0(\varphi)$ ($\varphi = \phi - \Theta$) satisfies the following equation:
\begin{equation}
  D \frac{d^2}{d\varphi^2} f_0\left( \varphi \right)
  + \left( \Omega - \omega \right) \frac{d}{d\varphi} f_0\left( \varphi \right)
  - \frac{d}{d\varphi} \Bigl[ g_0\left( \varphi \right) f_0\left( \varphi \right) \Bigr] = 0,
  \label{eq:f0-solution}
\end{equation}
where
\begin{equation}
  g_0\left( \varphi \right) = \int_{0}^{2\pi} d\varphi'\,
  \Gamma\left( \varphi - \varphi' \right) f_0\left( \varphi' \right).
\end{equation}
Let $u(\varphi, t)$ represent small disturbance to the collectively oscillating solution
and consider a slightly perturbed solution $f(\phi, t) = f_0(\varphi) + u(\varphi, t)$.
Equation~(\ref{eq:nfp}) with $\epsilon=0$ is linearized in $u(\varphi, t)$,
i.e., $\partial_t u(\varphi, t) = \hat{L} u(\varphi, t)$,
where the linear operator $\hat{L}$ is given by
\begin{equation}
  \hat{L} u\left( \varphi \right)
  = D \frac{d^2}{d\varphi^2} u\left( \varphi \right)
  + \left( \Omega - \omega \right) \frac{d}{d\varphi} u\left( \varphi \right)
  - \frac{d}{d\varphi} \Bigl[ g_0\left( \varphi \right) u\left( \varphi \right) \Bigr]
  - \frac{d}{d\varphi} \left[ f_0\left( \varphi \right) \int_{0}^{2\pi} d\varphi'\,
    \Gamma\left( \varphi - \varphi' \right) u\left( \varphi' \right) \right].
\end{equation}
Defining the inner product as
\begin{equation}
  \Bigl[ u^\ast\left( \varphi \right), u\left( \varphi \right) \Bigr]
  = \int_{0}^{2\pi} d\varphi\, u^\ast\left( \varphi \right) u\left( \varphi \right),
\end{equation}
we introduce an adjoint operator $\hat{L}^\ast$ of $\hat{L}$ by
\begin{equation}
  \left[ u^\ast\left( \varphi \right), \hat{L} u\left( \varphi \right) \right]
  = \left[ \hat{L}^\ast u^\ast\left( \varphi \right), u\left( \varphi \right) \right].
\end{equation}
The adjoint operator $\hat{L}^\ast$ is explicitly given as
\begin{equation}
  \hat{L}^{\ast} u^\ast\left( \varphi \right)
  = D \frac{d^2}{d\varphi^2} u^\ast\left( \varphi \right)
  - \left(\Omega-\omega\right) \frac{d}{d\varphi} u^\ast\left( \varphi \right)
  + g_0\left( \varphi \right) \frac{d}{d\varphi} u^\ast\left( \varphi \right)
  + \int_{0}^{2\pi} d\varphi'\, \Gamma\left( \varphi' - \varphi \right)
  f_0\left( \varphi' \right) \frac{d}{d\varphi'} u^\ast\left( \varphi' \right).
\end{equation}
In the calculation below, we need only zero eigenfunctions
$u_0(\varphi)$ of $\hat{L}$ and $u_0^\ast(\varphi)$ of $\hat{L}^\ast$.
Note that the right zero eigenfunction can be chosen as
\begin{equation}
  \hat{L} u_0\left( \varphi \right) = 0, \qquad
  u_0\left( \varphi \right) = \frac{d}{d\varphi} f_0\left( \varphi \right),
\end{equation}
which follows from differentiation of Eq.~(\ref{eq:f0-solution}) with respect to $\varphi$.
The left zero eigenfunction is normalized as
\begin{equation}
  \hat{L}^\ast u_0^\ast\left( \varphi \right) = 0, \qquad
  \left[ u_0^\ast\left( \varphi \right), u_0\left( \varphi \right) \right] = 1.
\end{equation}

Now let us introduce weak external coupling, i.e.,
we assume $0 < \epsilon \ll 1$ and treat the last term in Eq.~(\ref{eq:nfp}) as perturbations.
Using the phase reduction method~\cite{ref:kuramoto84,ref:kawamura07,ref:kawamura08},
we can derive coupled collective phase equations
from the nonlinear Fokker-Planck equations~(\ref{eq:nfp}).
Namely, we project the nonlinear Fokker-Planck equations~(\ref{eq:nfp})
onto the unperturbed collectively oscillating solution as
\begin{align}
  \frac{d}{dt}\left( - \Theta^{(\sigma)} \right)
  &= \left[ u_0^\ast\left( \phi - \Theta^{(\sigma)} \right),
    \frac{\partial}{\partial t} f^{(\sigma)}\left( \phi, t \right) \right] \nonumber \\
  &\simeq -\Omega - \epsilon \left[ u_0^\ast\left( \varphi \right),
    \frac{d}{d\varphi} \int_0^{2\pi} d\varphi'\,
    \Gamma_{\sigma\tau}\left( \varphi - \varphi'- \Theta^{(\sigma)} - \Theta^{(\tau)} \right)
    f_0\left( \varphi' \right) f_0\left( \varphi \right) \right],
\end{align}
where we approximated $f^{(\sigma)}(\phi,t)$ by the unperturbed solution $f_0(\varphi^{(\sigma)})$
and used that $[ u_0^{\ast}(\varphi), \dot{f_{0}}(\varphi) ] = -\Omega$.
Therefore, the collective phase equation takes the form
\begin{equation}
  \dot{\Theta}^{(\sigma)} = \Omega
  + \epsilon \, \digamma_{\sigma\tau}\left( \Theta^{(\sigma)} - \Theta^{(\tau)} \right),
  \label{eq:collective}
\end{equation}
where the {\it collective phase coupling function} is given by
\begin{equation}
  \digamma_{\sigma\tau}\left( \Theta^{(\sigma)} - \Theta^{(\tau)} \right)
  = \int_{0}^{2\pi} d\varphi \int_{0}^{2\pi} d\varphi' \,
  \Gamma_{\sigma\tau}\left( \varphi - \varphi' + \Theta^{(\sigma)} - \Theta^{(\tau)} \right)
  k_0\left( \varphi \right) f_0\left( \varphi' \right)
  \label{eq:digamma}
\end{equation}
for $(\sigma, \tau) = (1, 2)$ or $(2, 1)$.
The function $k_0(\varphi)$ is defined by
\begin{equation}
  k_0\left( \varphi \right)
  = -f_0\left( \varphi \right) \frac{d}{d\varphi} u_0^\ast\left( \varphi \right)
\end{equation}
and normalized as
\begin{equation}
  \int_{0}^{2\pi} d\varphi \, k_0\left( \varphi \right)
  = \int_{0}^{2\pi} d\varphi \, u_0^\ast\left( \varphi \right) u_0\left( \varphi \right) = 1,
\end{equation}
which is the kernel function determining the {\it collective phase sensitivity} of the group
as a convolution of the microscopic phase sensitivity~\cite{ref:kawamura08}.

\subsection{The case with sinusoidal coupling}

When the microscopic external phase coupling function $\Gamma_{\sigma\tau}(\phi)$ is sinusoidal as given in Eq.~(\ref{eq:beta}),
the collective phase coupling function also takes a sinusoidal form
\begin{equation}
  \digamma_{\sigma\tau}\left( \Theta \right) = - \rho \sin\left( \Theta + \delta \right)
  \label{eq:sinusoidal}
\end{equation}
because Eq.~(\ref{eq:digamma}) is a double convolution of $\Gamma_{\sigma\tau}(\varphi - \varphi' + \Theta)$
with $k_0(\varphi) f_0(\varphi')$.
Here, the parameter $\rho\cos\delta$ determines the effective type of the collective phase coupling function;
it is in-phase when $\rho \cos \delta > 0$ and anti-phase when $\rho \cos \delta < 0$.
This quantity can be evaluated from the following equation:
\begin{equation}
  \rho \cos \delta
  = -\left. \frac{d \digamma_{\sigma\tau}\left( \Theta \right)}{d\Theta} \right|_{\Theta=0}
  = -\int_{0}^{2\pi} d\varphi \int_{0}^{2\pi} d\varphi' \,
  \Gamma_{\sigma\tau}\left( \varphi - \varphi' \right)
  k_0\left( \varphi \right) u_0\left( \varphi' \right).
  \label{eq:type}
\end{equation}

\begin{figure*}
  \begin{center}
    \includegraphics[width=0.45\hsize,clip]{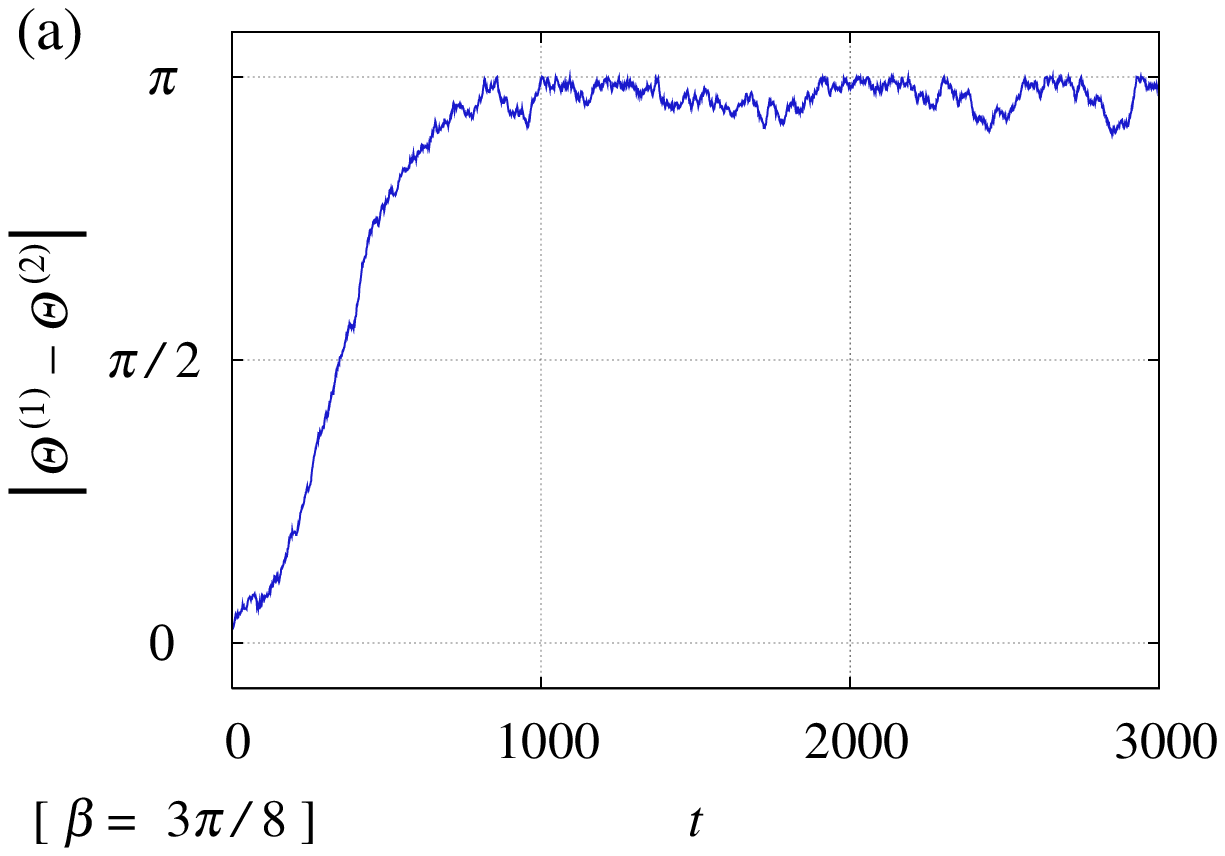}
    \includegraphics[width=0.45\hsize,clip]{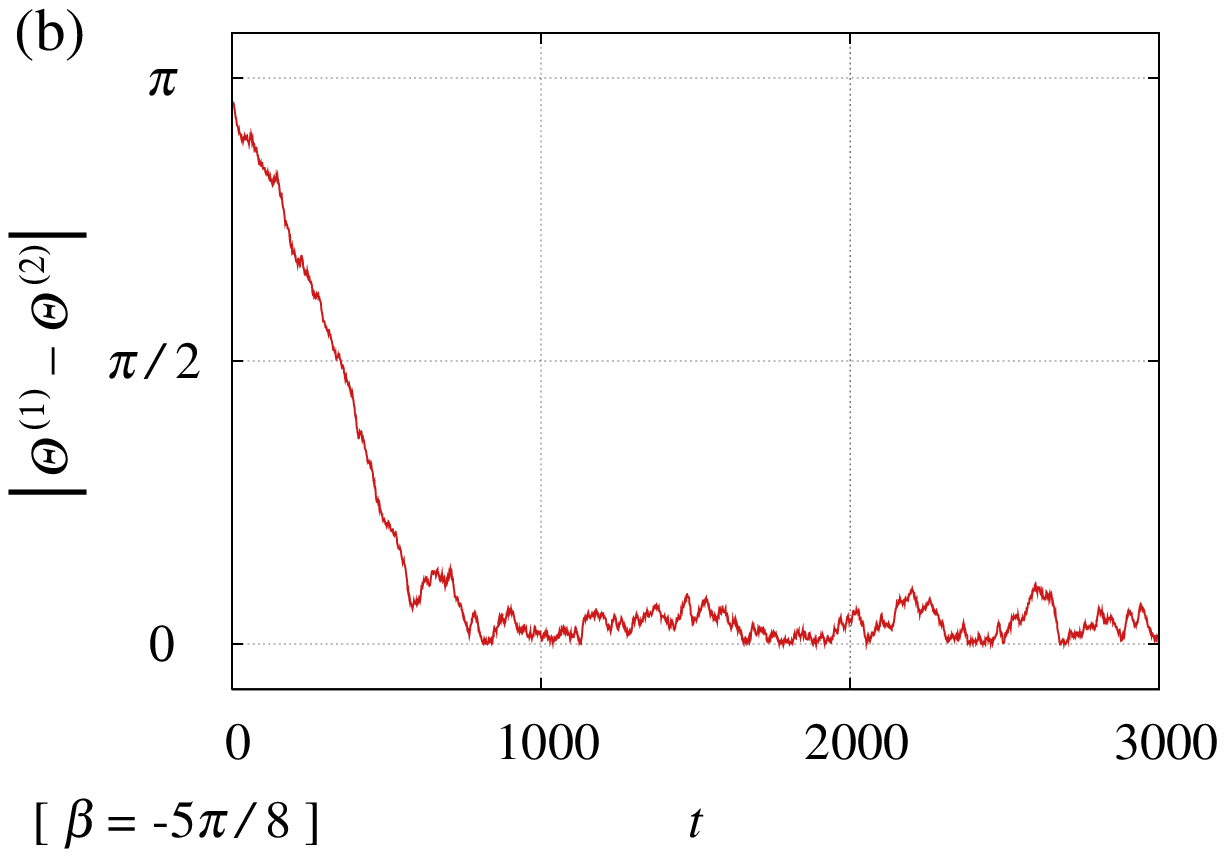}
    \caption{(Color online)
      Time evolution of collective phase difference
      $| \Theta^{(1)} - \Theta^{(2)} |$.
      (a) Effective anti-phase collective synchronization
      with microscopic in-phase external coupling, $\beta = 3 \pi / 8$.
      (b) Effective in-phase collective synchronization
      with microscopic anti-phase external coupling, $\beta = -5 \pi / 8$.
      The other parameters are $\epsilon = 0.01$,
      $\omega = 0$,
      $D = D_{\rm c} / 2 = (\cos \alpha) / 4$,
      and $\alpha = 3 \pi / 8$.
      In numerical simulations,
      the number of oscillator in each group is $N = 10000$.
    }
    \label{fig:1}
  \end{center}
\end{figure*}

\begin{figure*}
  \begin{center}
    \includegraphics[width=0.45\hsize,clip]{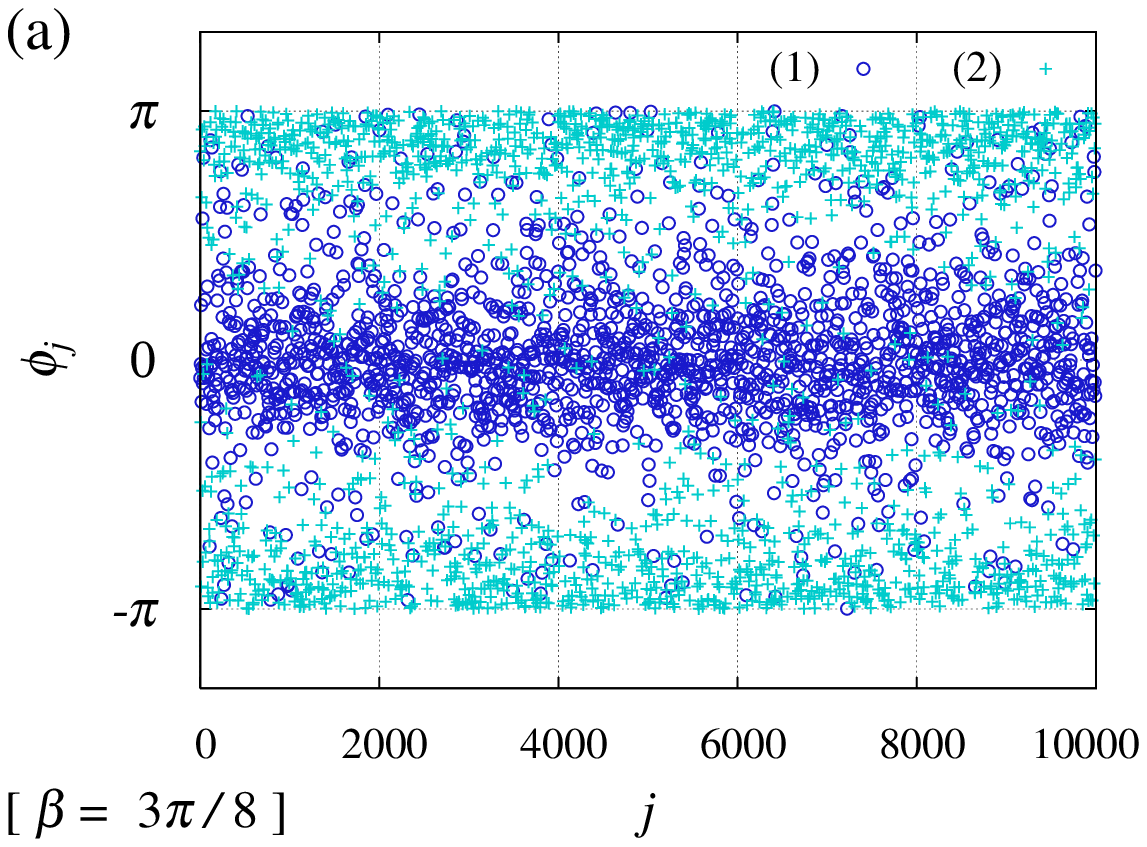}
    \includegraphics[width=0.45\hsize,clip]{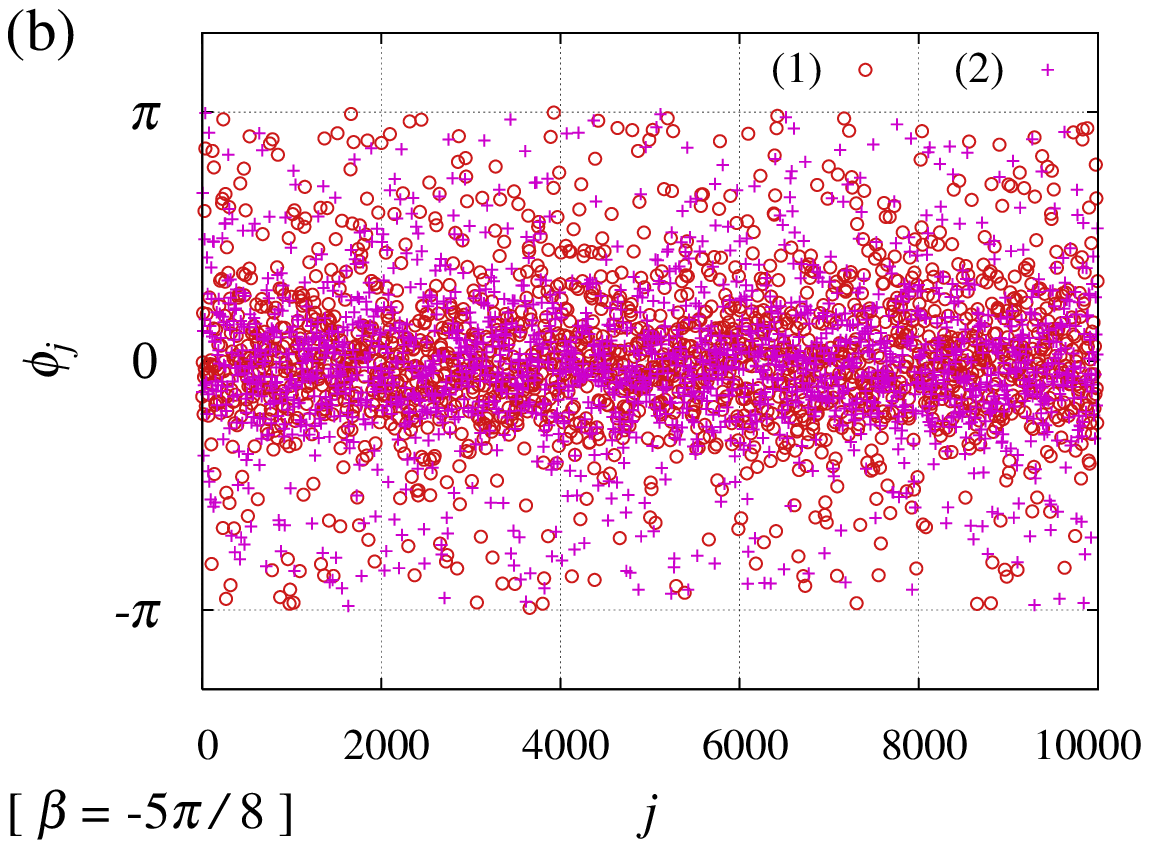}
    \caption{(Color online)
      Snapshots of the asymptotic states of the individual oscillators in Fig.~\ref{fig:1}.
      Only one in every five oscillators is plotted.
      Open circle ($\circ$) and plus ($+$) respectively indicate
      oscillator of group~$(1)$ and that of group~$(2)$.
      (a) Effective anti-phase coupling
      with microscopic in-phase coupling, $\beta = 3 \pi / 8$.
      (b) Effective in-phase coupling
      with microscopic anti-phase coupling, $\beta = -5 \pi / 8$.
    }
    \label{fig:2}
  \end{center}
\end{figure*}

\begin{figure*}
  \begin{center}
    \begin{tabular}{c c}
      \includegraphics[width=0.45\hsize,clip]{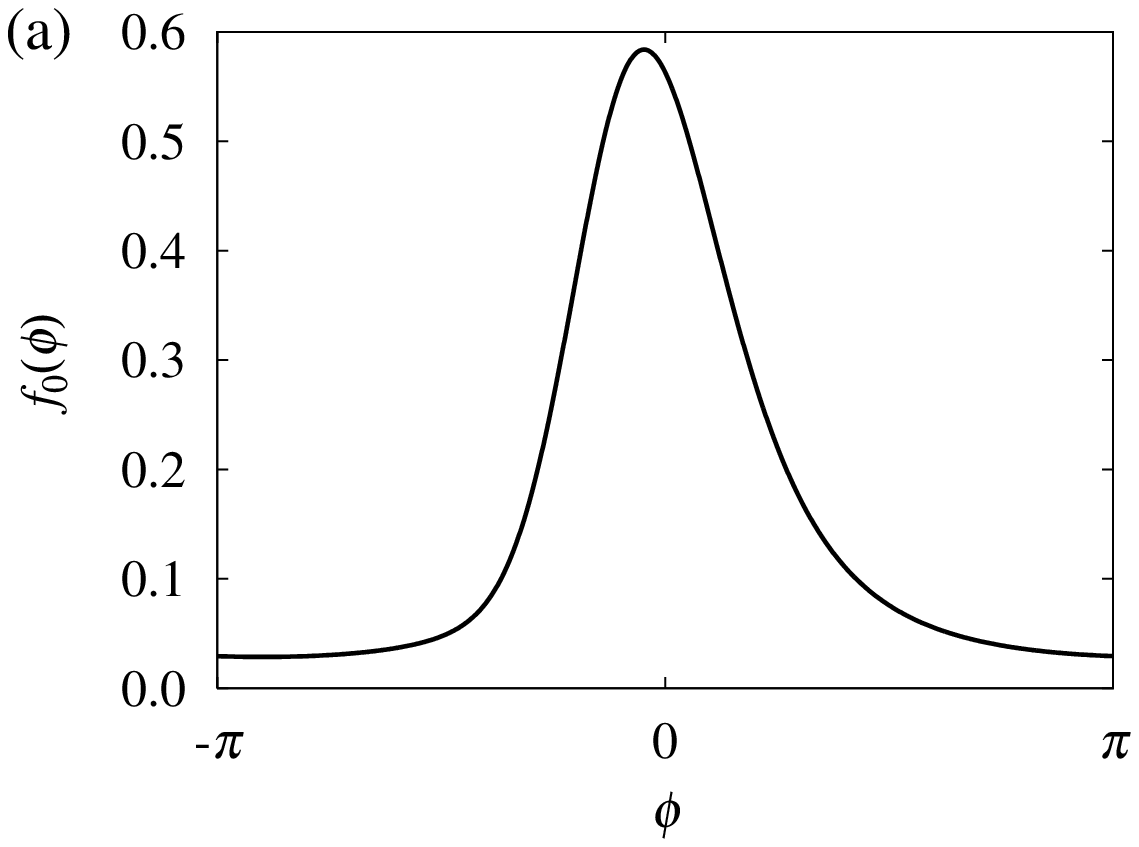} &
      \includegraphics[width=0.45\hsize,clip]{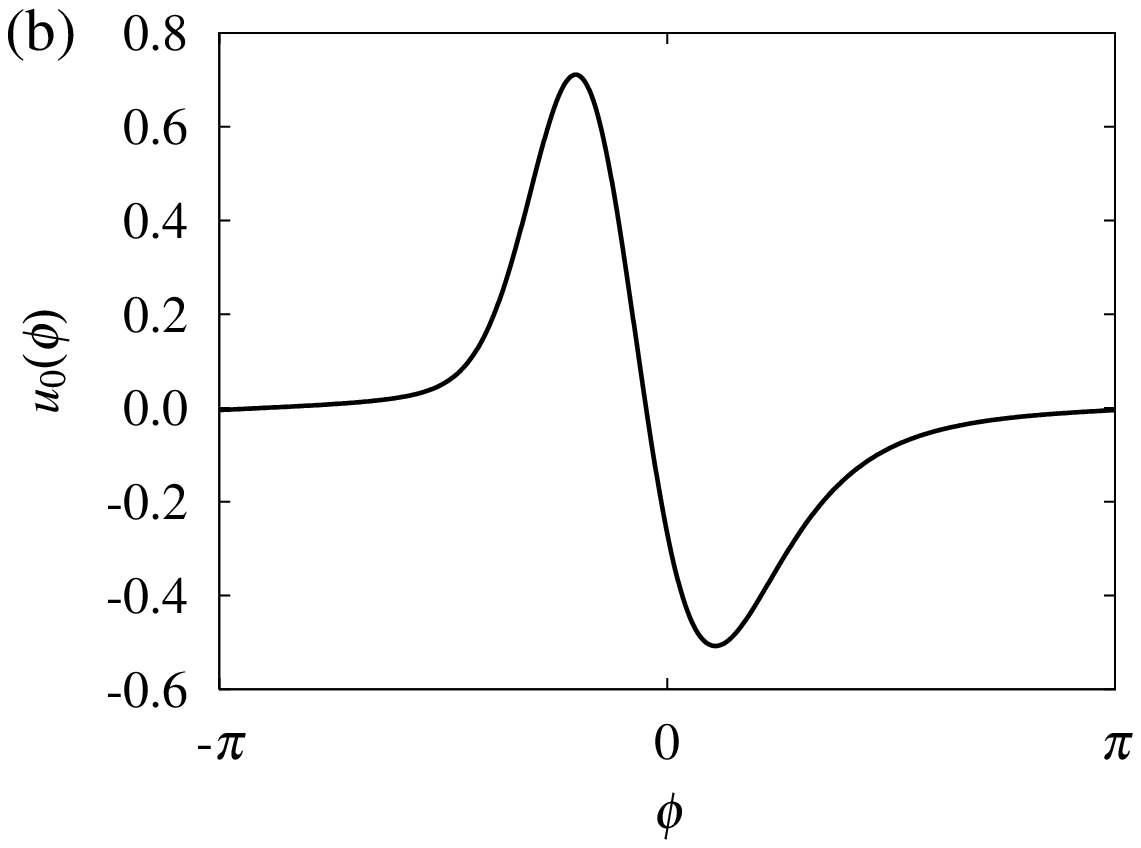} \\
      \includegraphics[width=0.45\hsize,clip]{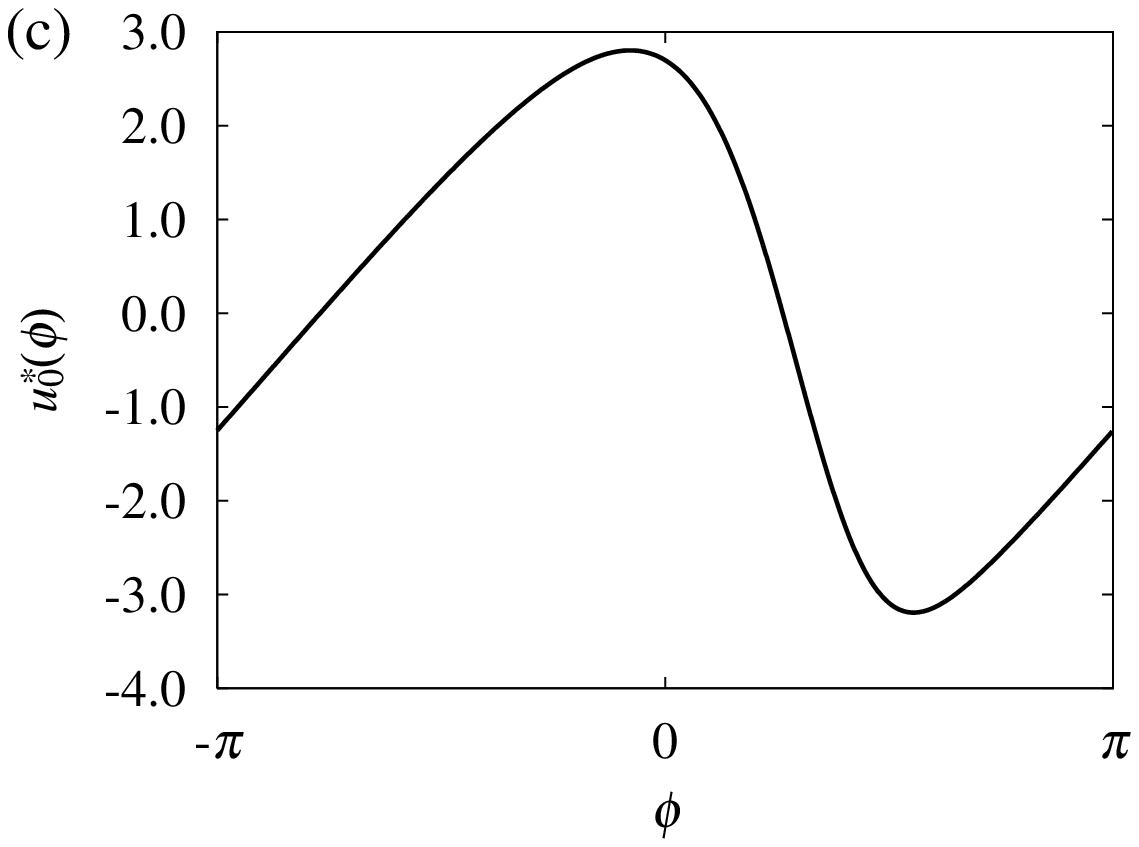} &
      \includegraphics[width=0.45\hsize,clip]{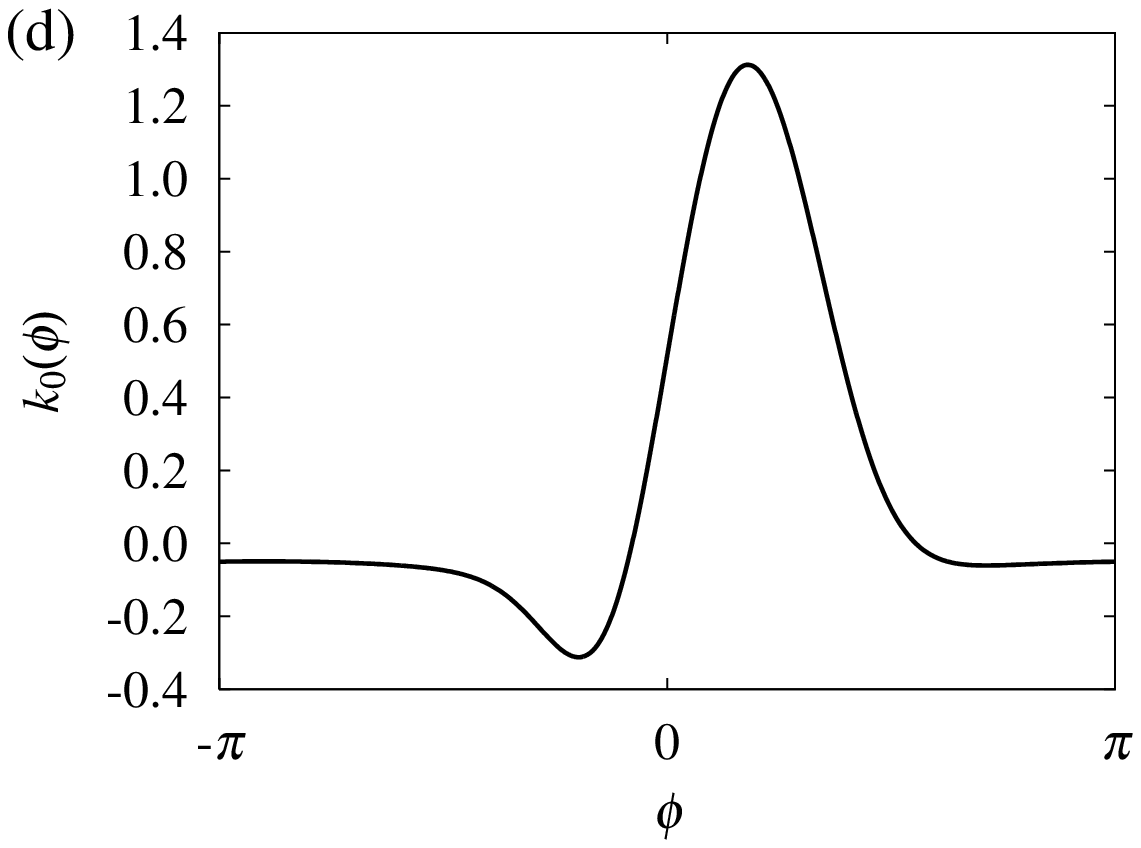} \\
    \end{tabular}
    \caption{
      (a) Distribution function $f_0(\phi)$.
      (b) Right zero eigenfunction $u_0(\phi)$.
      (c) Left zero eigenfunction $u_0^\ast(\phi)$.
      (d) Kernel function $k_0(\phi)$.
      Parameters are $D = D_{\rm c}/2 = (\cos\alpha)/4$ and $\alpha = 3\pi/8$,
      where order parameter amplitude is $R = 0.65328$.
      $\beta =  3\pi/8$ gives $\rho \cos \delta = -0.32264$, whereas
      $\beta = -5\pi/8$ gives $\rho \cos \delta =  0.32264$.
    }
    \label{fig:3}
  \end{center}
\end{figure*}

\begin{figure*}
  \begin{center}
    \includegraphics[width=0.50\hsize,clip]{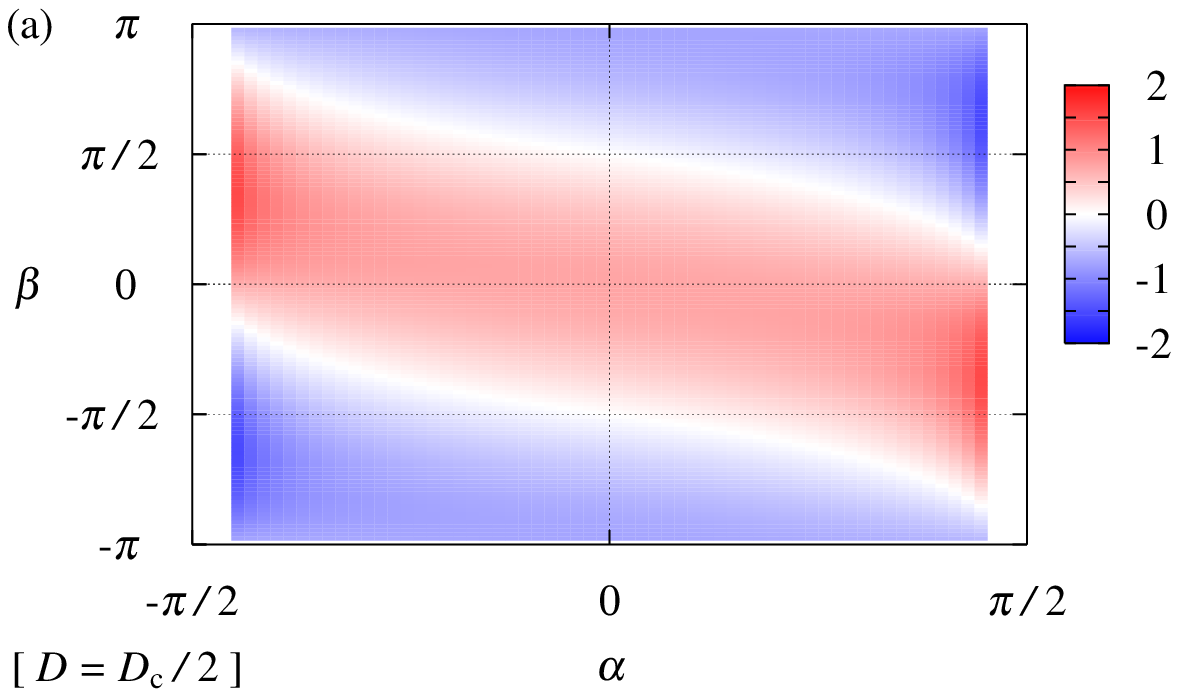}
    \includegraphics[width=0.42\hsize,clip]{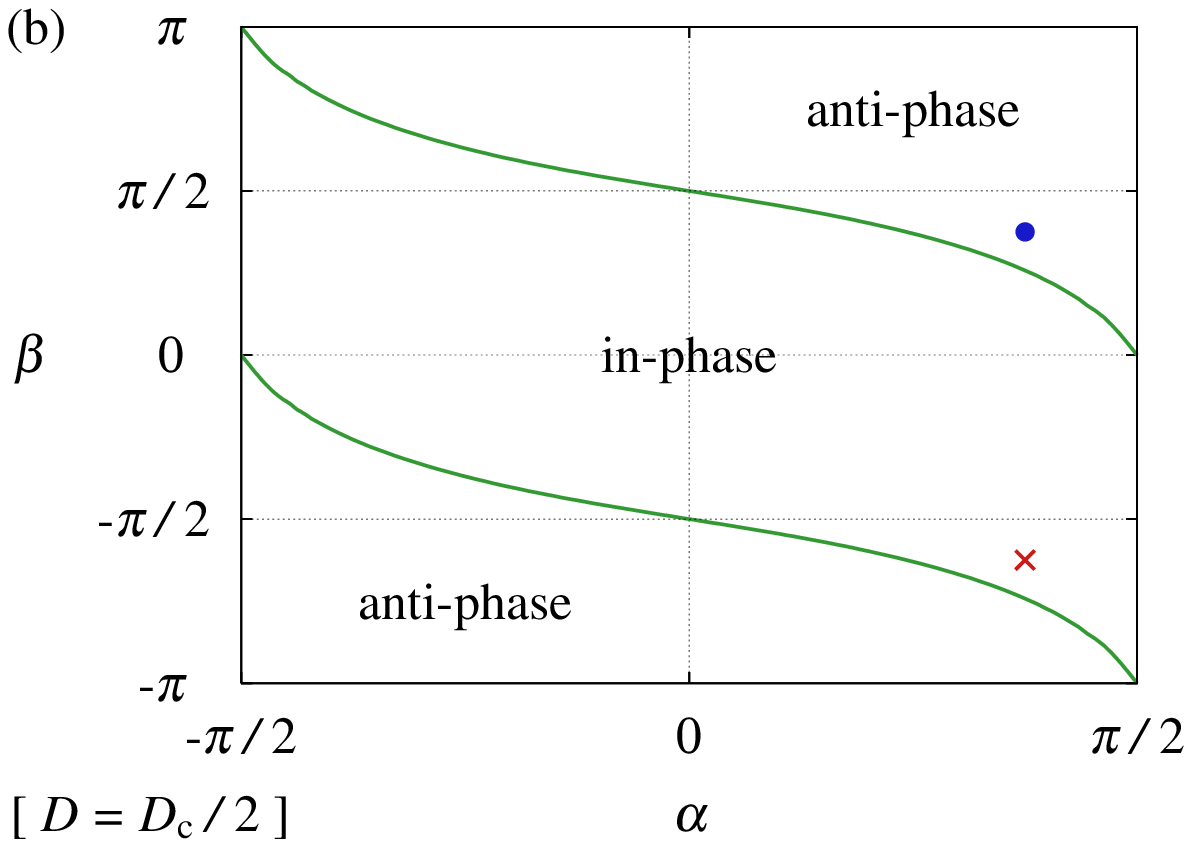}
    \caption{(Color online)
      Effective type of phase coupling
      between collective oscillations with
      $\alpha \in ( -\pi / 2, \pi / 2 )$,
      $\beta \in [ -\pi, \pi ]$, and
      $D = D_{\rm c} / 2 = (\cos \alpha) / 4$,
      which is numerically evaluated by Eq.~(\ref{eq:type}).
      (a) Dependence of $\rho \cos \delta$
      on $\alpha$ and $\beta$.
      (b) The solid curves are determined by
      $\rho \cos \delta = 0$.
      The filled circle ($\bullet$) indicates
      $\alpha = \beta = 3 \pi / 8$
      corresponding to Figs.~\ref{fig:1}(a) and \ref{fig:2}(a).
      The cross ($\times$) indicates
      $\alpha = 3 \pi / 8$ and $\beta = -5 \pi / 8$
      corresponding to Figs.~\ref{fig:1}(b) and \ref{fig:2}(b).
    }
    \label{fig:4}
  \end{center}
\end{figure*}

Now we examine the case $D = D_{\rm c} / 2 = (\cos \alpha) / 4$,
which we considered in the numerical simulations shown in Fig.~\ref{fig:1}.
Typical functional shapes of $f_0(\phi)$, $u_0(\phi)$, $u_0^\ast(\phi)$, and $k_0(\phi)$ in this case
are illustrated in Fig.~\ref{fig:3}, which were numerically obtained from the nonlinear Fokker-Planck equation.
Details of the numerical method are described in Ref.~\cite{ref:kawamura07}.
From these functions, the dependence of $\rho \cos \delta$ on $\alpha$ and $\beta$
was numerically evaluated by Eq.~(\ref{eq:type}) as shown in Fig.~\ref{fig:4}(a).
The type of the collective phase coupling function is represented in Fig.~\ref{fig:4}(b),
where the solid curves satisfying $\rho \cos \delta = 0$
represent the borders between the in-phase and the anti-phase parameter regions.
The two sets of parameter values used in Fig.~\ref{fig:1} are also plotted in Fig.~\ref{fig:4}(b).
As can be seen, the set of parameters corresponding to Fig.~\ref{fig:1}(a)
is in the anti-phase region, $\rho \cos \delta < 0$,
which yields effective anti-phase collective phase coupling between the groups.
Similarly, the parameter set corresponding to Fig.~\ref{fig:1}(b)
is in the in-phase region, $\rho \cos \delta > 0$,
yielding effective in-phase collective phase coupling.
Thus, the collective phase reduction theory successfully explains the numerical results in Fig.~\ref{fig:1}.

\section{Center-manifold and phase reductions} \label{sec:amplitude}

In this section, we analytically determine the collective phase coupling function
at the onset of collective oscillations by applying phase reduction to amplitude equations
obtained by the center-manifold reduction of the nonlinear Fokker-Planck equations.
This method gives analytical results without recourse to numerical determination
of the kernel and other functions for general microscopic phase coupling functions,
though restricted to the vicinity of the onset of collective oscillations.

\subsection{Amplitude equations near the onset of collective oscillations}

We derive coupled amplitude equations that describe the macroscopic rhythms of the groups
near the onset of collective oscillations.
Expanding the $2\pi$-periodic functions $f^{(\sigma)}(\phi, t)$, $\Gamma(\phi)$, and $\Gamma_{\sigma\tau}(\phi)$
into Fourier series as
\begin{equation}
  f^{(\sigma)}\left( \phi, t \right) =
  \frac{1}{2\pi} \sum_{l=-\infty}^{\infty}
  f_l^{(\sigma)}\left( t \right) e^{il\phi}, \qquad
  f_l^{(\sigma)}\left( t \right) = \int_{0}^{2\pi} d\phi\,
  f^{(\sigma)}\left( \phi, t \right) e^{-il\phi},
\end{equation}
\begin{equation}
  \Gamma\left( \phi \right) = \sum_{l=-\infty}^{\infty} \Gamma_l\, e^{il\phi}, \qquad
  \Gamma_l = \frac{1}{2\pi} \int_{0}^{2\pi} d\phi\, \Gamma\left( \phi \right) e^{-il\phi},
\end{equation}
\begin{equation}
  \Gamma_{\sigma\tau}\left( \phi \right) =
  \sum_{l=-\infty}^{\infty} \Gamma_{\sigma\tau, l}\, e^{il\phi}, \qquad
  \Gamma_{\sigma\tau, l} = \frac{1}{2\pi} \int_{0}^{2\pi} d\phi\,
  \Gamma_{\sigma\tau}\left( \phi \right) e^{-il\phi},
\end{equation}
the coupled nonlinear Fokker-Planck equations~(\ref{eq:nfp}) can be expressed as
\begin{align}
  \dot{f}_l^{(\sigma)}(t) =
  &- \left[ D l^2 + i l \left( \omega + \Gamma_0 + \Gamma_l \right) \right] f_l^{(\sigma)}
  - i l \sum_{m \ne 0, l} \Gamma_m f_m^{(\sigma)} f_{l-m}^{(\sigma)} \nonumber \\
  &+ \epsilon \left[ - i l \Gamma_{\sigma\tau, 0} f_l^{(\sigma)}
    - i l \Gamma_{\sigma\tau, l} f_l^{(\tau)}
    - i l \sum_{m \ne 0, l} \Gamma_{\sigma\tau, m} f_{m}^{(\tau)} f_{l-m}^{(\sigma)} \right].
  \label{eq:nfp2}
\end{align}

When the noise intensity $D$ is decreased below the critical value $D_{\rm c}$
in the absence of external coupling between the groups, $\epsilon = 0$,
the uniform solution $f^{(\sigma)}(\phi, t) = 1 / (2 \pi)$ of Eq.~(\ref{eq:nfp}),
corresponding to the incoherent state, becomes unstable.
Equivalently, the trivial solution $f_0^{(\sigma)} = 1$, $f_l^{(\sigma)} = 0 \; (l \neq 0)$
of Eq.~(\ref{eq:nfp2}) is destabilized and a pair of modes $f_{\pm l_{\rm c}}^{(\sigma)}$
with critical non-zero wavenumbers $\pm l_{\rm c}$ start to grow.
From the linear part of Eq.~(\ref{eq:nfp2}),
instability of the mode $l$ occurs when $D l^{2} - l {\rm Im} \Gamma_{l} < 0$,
namely, $D < {\rm Im} \Gamma_{l} / l$.
Thus, the most unstable wavenumbers $\pm l_{\rm c}$ are those that maximize ${\rm Im} \Gamma_{l} / l$.
Generally, the fundamental harmonic components tend to be predominant in the phase coupling function,
so that we obtain $l_{\rm c} = \pm 1$ in most cases.

Assuming $l_{\rm c} = \pm 1$,
we introduce a complex amplitude $A^{(\sigma)}(t)$
of the fundamental harmonic modes of $f^{(\sigma)}(\phi, t)$ as
\begin{equation}
  f^{(\sigma)}\left( \phi, t \right) = \frac{1}{2\pi}
  + \frac{1}{2\pi} \Bigl( A^{(\sigma)}\left( t \right) e^{-i\phi}
  + \bar{A}^{(\sigma)}\left( t \right) e^{i\phi} \Bigr),
\end{equation}
where $\bar{A}^{(\sigma)}(t)$ is the complex conjugate of
$A^{(\sigma)}(t) = f^{(\sigma)}_{-1}(t)$.
We are concerned with weakly coupled collective oscillations,
and thus consider the external interaction as perturbations.
Using the center-manifold reduction method~\cite{ref:kuramoto84},
we can derive a pair of coupled complex amplitude equations
from Eq.~(\ref{eq:nfp2}) in the following form~\cite{ref:okuda91}:
\begin{equation}
  \dot{A}^{(\sigma)}
  = \left( \mu + i \Omega_{\rm c} \right) A^{(\sigma)}
  - g \left| A^{(\sigma)} \right|^2 A^{(\sigma)}
  + \epsilon d_{\sigma\tau} A^{(\tau)}
  \label{eq:amplitude}
\end{equation}
for $(\sigma, \tau) = (1,2)$ or $(2,1)$, where the parameters are given by
\begin{equation}
  \mu = D_{\rm c} - D, \qquad
  D_{\rm c} = - {\rm Im} \Gamma_{-1}, \qquad
  \Omega_{\rm c} = \omega + {\rm Re} \Gamma_{-1} + \Gamma_0 + \epsilon \Gamma_{\sigma\tau, 0},
  \label{eq:parameter1}
\end{equation}
and
\begin{equation}
  g = \frac{- \Gamma_{-1} \left( \Gamma_{-2} + \Gamma_{1} \right)}
  {2 {\rm Im} \Gamma_{-1} - i {\rm Re} \Gamma_{-1} + i \Gamma_{-2}}, \qquad
  d_{\sigma\tau} = i\, \Gamma_{\sigma\tau, -1}.
  \label{eq:parameter2}
\end{equation}
See Refs.~\cite{ref:kuramoto84,ref:okuda91,ref:kawamura07} for details of the derivation.
We should note that Eq.~(\ref{eq:amplitude}) represents two coupled Stuart-Landau oscillators,
each of which (i.e., $\dot{A} = ( \mu + i \Omega_{\rm c} ) A - g |A|^2 A$)
describes collective oscillations of the respective oscillator group.

\subsection{Phase reduction of the amplitude equations}

Next, we derive coupled collective phase equations by reducing the coupled Stuart-Landau equations obtained above
by assuming that the external interaction between the groups is sufficiently weak, i.e., $\epsilon$ is small.
When the two groups are uncoupled, $\epsilon = 0$,
the limit-cycle solution $A_0(\Theta)$ of Eq.~(\ref{eq:amplitude}) is given by
($\sigma$ is dropped again for the moment)
\begin{equation}
  A_0(\Theta) = \sqrt{\frac{\mu}{{\rm Re}\, g}} e^{i \Theta}, \qquad
  \dot{\Theta} = \Omega = \Omega_{\rm c} - \mu \frac{{\rm Im}\, g}{{\rm Re}\, g}.
  \label{eq:solution}
\end{equation}
The left and right Floquet eigenvectors of this limit-cycle solution associated with the zero eigenvalue
can be written as
\begin{equation}
  U_0(\Theta) = \frac{dA_0(\Theta)}{d\Theta} = i \sqrt{\frac{\mu}{{\rm Re}\, g}} e^{i \Theta}, \qquad
  U_0^\ast(\Theta) = i \sqrt{\frac{{\rm Re}\, g}{\mu}} \frac{g}{{\rm Re}\, g} e^{i \Theta},
  \label{eq:eigenvector}
\end{equation}
where the inner product of $U_{0}(\Theta)$ and $U_0^\ast(\Theta)$ satisfies the normalization condition
\begin{equation}
  {\rm Re}\, \Bigl[ \bar{U}_0^\ast\left( \Theta \right) U_0\left( \Theta \right) \Bigr] = 1.
  \label{eq:normalization}
\end{equation}
Though Eq.~(\ref{eq:eigenvector}) is expressed in complex representation for the sake of convenience
in analytical calculations performed below, they are equivalent to the known results~\cite{ref:kuramoto84}.

Now let us introduce weak external coupling as perturbations, i.e., we assume $0 < \epsilon \ll \mu \ll 1$.
Using the phase reduction method~\cite{ref:kuramoto84},
we can derive the collective phase equation~(\ref{eq:collective})
from the amplitude equation~(\ref{eq:amplitude}).
Namely, we project the amplitude equation~(\ref{eq:amplitude}) onto the unperturbed limit-cycle orbit as
\begin{equation}
  \dot{\Theta}^{(\sigma)}
  = {\rm Re} \left[ \bar{U}_0^\ast\left( \Theta^{(\sigma)} \right) \dot{A}^{(\sigma)} \right]
  \simeq \Omega
  + \epsilon {\rm Re} \left[ \bar{U}_0^\ast\left( \Theta^{(\sigma)} \right) d_{\sigma \tau} A_0\left( \Theta^{(\tau)} \right) \right],
\end{equation}
where we approximated $A^{(\sigma)}$ by the unperturbed solution $A_0(\Theta^{(\sigma)})$
and used that ${\rm Re} [ \bar{U}_0^\ast(\Theta) \dot{A_0}(\Theta) ] = \Omega$.
Thus, the reduced equation is obtained in the form of Eq.~(\ref{eq:collective}),
and the collective phase coupling function is given by
\begin{equation}
  \digamma_{\sigma\tau}\left( \Theta^{(\sigma)} - \Theta^{(\tau)} \right)
  = {\rm Re} \left[ \bar{U}_0^\ast\left( \Theta^{(\sigma)} \right)
    d_{\sigma\tau} A_0\left( \Theta^{(\tau)} \right) \right].
  \label{eq:digamma1}
\end{equation}
By inserting the expressions of Eqs.~(\ref{eq:parameter1}), (\ref{eq:parameter2}), (\ref{eq:solution}), and (\ref{eq:eigenvector})
into the formula Eq.~(\ref{eq:digamma1}),
the collective phase coupling function $\digamma_{\sigma\tau}(\Theta)$ can be analytically obtained,
which takes a sinusoidal form
\begin{equation}
  \digamma_{\sigma\tau}\left( \Theta \right)
  = -\rho \sin\left( \Theta  + \delta \right), \qquad
  \rho e^{i \delta} = \frac{g \bar{d}_{\sigma\tau}}{{\rm Re}\, g}.
  \label{eq:digamma2}
\end{equation}
Note that we have not assumed that the external phase coupling function $\Gamma_{\sigma\tau}(\phi)$ is sinusoidal so far.
The sinusoidal collective phase coupling function arises
because we assume that collective oscillations
exhibited by the groups of oscillators are near the supercritical Hopf bifurcation point.

\begin{figure*}
  \begin{center}
    \includegraphics[width=0.50\hsize,clip]{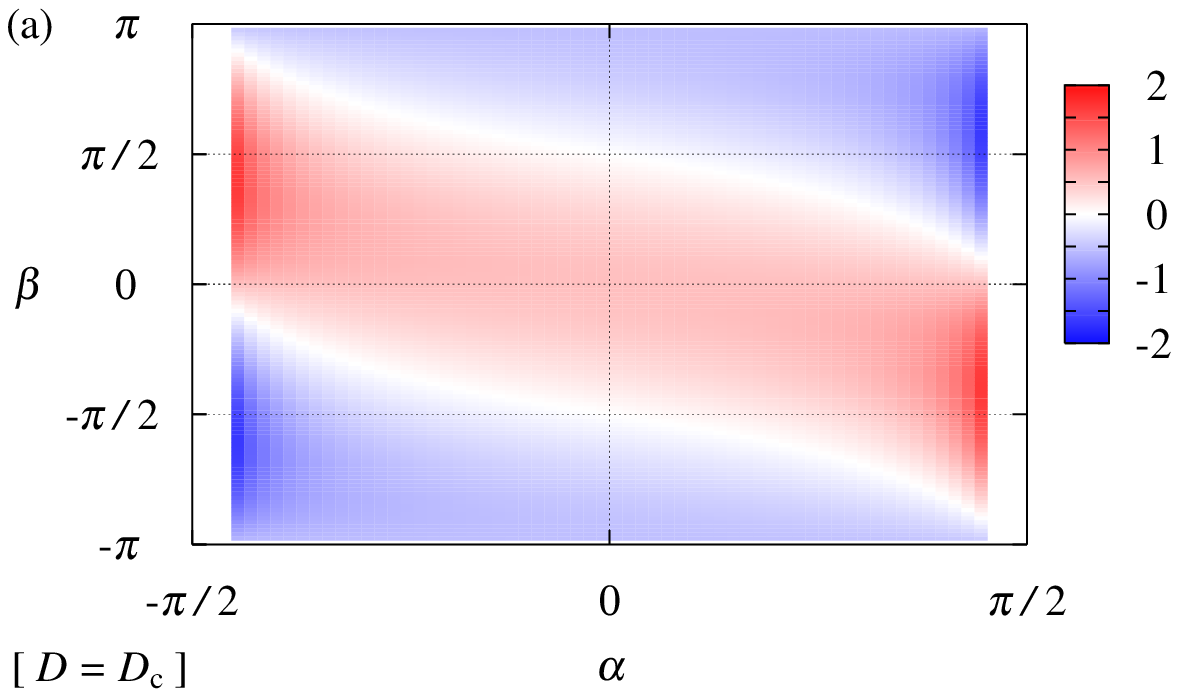}
    \includegraphics[width=0.42\hsize,clip]{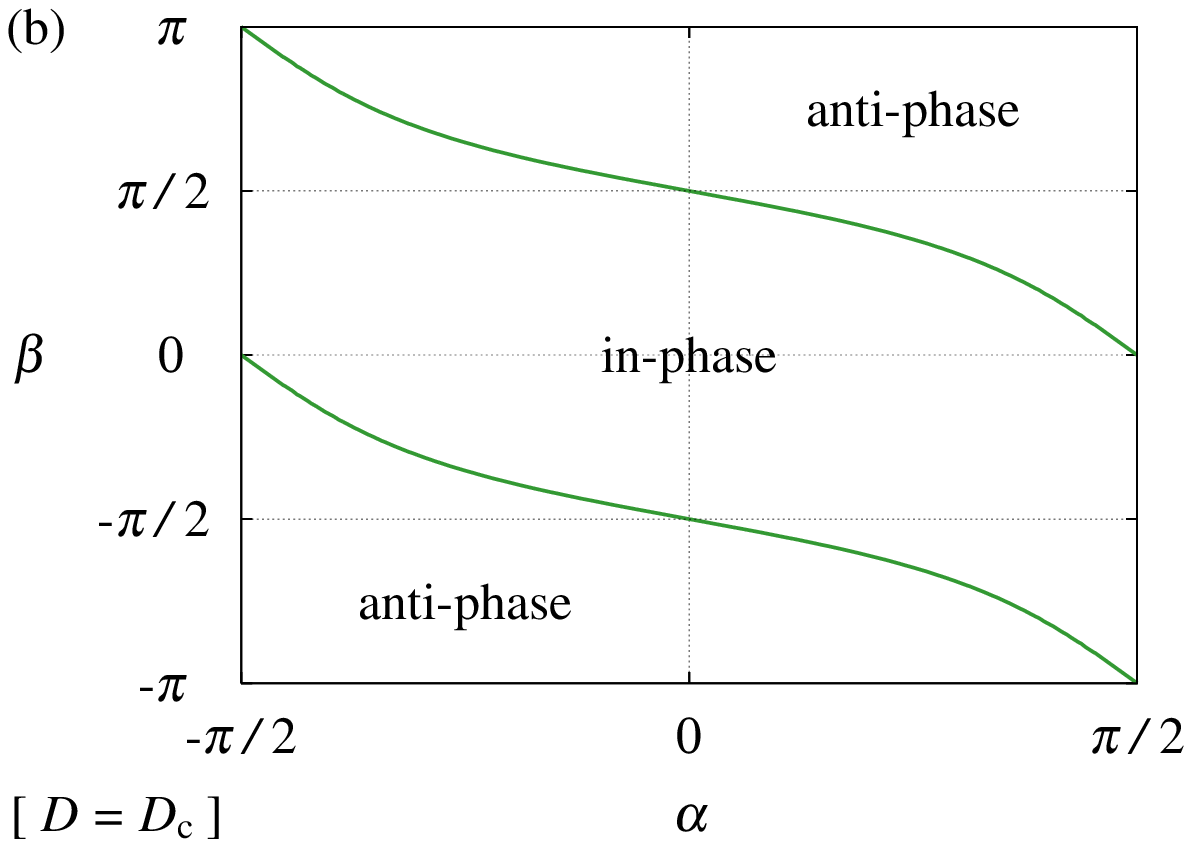}
    \caption{(Color online)
      Effective type of phase coupling between collective oscillations
      in $\alpha$ and $\beta$ with $D = D_{\rm c} = (\cos \alpha) / 2$,
      which is analytically given by Eq.~(\ref{eq:type2}), i.e.,
      $\rho \cos \delta = (2 \cos \beta - \tan \alpha \sin \beta) / 4$.
      (a) Dependence of $\rho \cos \delta$ on $\alpha$ and $\beta$.
      (b) The solid curves are determined by $\rho \cos \delta = 0$.
    }
    \label{fig:5}
  \end{center}
\end{figure*}

When the phase coupling functions are given by the sinusoidal forms, Eqs.~(\ref{eq:alpha}) and (\ref{eq:beta}),
the parameters of Eqs.~(\ref{eq:parameter1}), (\ref{eq:parameter2}), and (\ref{eq:solution}) can be calculated as
\begin{equation}
  D_{\rm c} = \frac{\cos \alpha}{2}, \qquad
  \Omega_{\rm c} = \omega - \frac{\sin \alpha}{2}, \qquad
  \Omega = \omega - \frac{3 \sin \alpha}{4} + \frac{D \tan \alpha}{2},
  \label{eq:parameter1sin}
\end{equation}
and
\begin{equation}
  g = \frac{1}{4\cos\alpha - 2i\sin\alpha}, \qquad
  d_{\sigma\tau} = \frac{1}{2} e^{-i\beta}.
  \label{eq:parameter2sin}
\end{equation}
Inserting Eq.~(\ref{eq:parameter2sin}) into Eq.~(\ref{eq:digamma2}), we obtain
\begin{equation}
  \rho e^{i \delta}
  = \frac{1}{4} \bigl( 2\cos\beta - \tan\alpha \sin\beta \bigr)
  + \frac{i}{4} \bigl( 2\sin\beta + \tan\alpha \cos\beta \bigr).
  \label{eq:type1}
\end{equation}
Therefore, the type of the collective phase coupling function is analytically found from the following quantity:
\begin{equation}
  \rho \cos \delta = \frac{1}{4} \bigl( 2\cos\beta - \tan\alpha \sin\beta \bigr).
  \label{eq:type2}
\end{equation}
Reflecting the symmetry of the original model of Eq.~(\ref{eq:simulation})
with respect to $(\alpha, \beta) \to -(\alpha, \beta)$ and $\phi \to -\phi$,
Eq.~(\ref{eq:type2}) is symmetric about the origin in the $\alpha$-$\beta$ plane.
The type of the collective phase coupling function
at the onset of collective oscillations, i.e., $D = D_{\rm c}$,
is represented in Fig.~\ref{fig:5}, which is very similar to Fig.~\ref{fig:4}.

\subsection{Several important cases}

Here, we consider three special and important cases of the collective phase coupling functions
derived for the sinusoidal microscopic phase coupling functions, Eqs.~(\ref{eq:alpha}) and (\ref{eq:beta}),
at the onset of collective oscillations.

(i) The first case is $\alpha = 0$,
which indicates that the internal phase coupling function within the same group is antisymmetric.
Inserting $\alpha = 0$ into Eq.~(\ref{eq:type1}), we obtain the following result:
\begin{equation}
  \rho e^{i \delta} = \frac{1}{2} e^{i \beta},
\end{equation}
so that $\rho \cos \delta = ( \cos \beta ) / 2$.
Thus, the collective phase coupling function has the same type as the microscopic external phase coupling function.
The internal phase coupling function does not affect the type of the collective phase coupling.
A similar scenario has been encountered in different models~\cite{ref:kawamura08,ref:kori09,ref:kawamura10}.

(ii) Several special values of the microscopic external coupling phase shift $\beta$ comprise the second case.
Inserting $\beta = 0, \pm\pi, \pm\pi/2$ into Eq.~(\ref{eq:type1}),
we obtain the following results:
\begin{equation}
  \beta = 0, \qquad
  \rho e^{i \delta} = \frac{1}{2} + \frac{i}{4} \tan\alpha,
\end{equation}
\begin{equation}
  \beta = \pm\pi, \qquad
  \rho e^{i \delta} = -\frac{1}{2} - \frac{i}{4} \tan\alpha,
\end{equation}
\begin{equation}
  \beta = \pm\frac{\pi}{2}, \qquad
  \rho e^{i \delta} = \mp\frac{1}{4}\tan\alpha \pm \frac{i}{2}.
\end{equation}
For antisymmetric external interactions, i.e., $\beta = 0, \pm\pi$,
the type of the collective phase coupling function coincides with the microscopic external coupling
and is not affected by the type of the microscopic internal coupling phase shift $\alpha$.
In contrast, for symmetric external interactions, i.e., $\beta = \pm\pi/2$,
the type of the collective phase coupling function
is solely determined by the internal coupling parameter $\alpha$,
which can be either in-phase or anti-phase.

(iii) The third case is $\beta = \alpha$, namely, when the external coupling has the same phase shift as the internal one.
Inserting $\beta = \alpha$ into Eq.~(\ref{eq:type1}),
we obtain the following result:
\begin{equation}
  \rho e^{i \delta}
  = \frac{1}{4} \bigl( 2\cos\alpha - \tan\alpha \sin\alpha \bigr)
  + i \frac{3\sin\alpha}{4}.
\end{equation}
Note that $| \beta | = | \alpha | < \pi / 2$ in this case, namely,
both internal and external coupling functions are in-phase.
The type of the collective phase coupling function is anti-phase when $\tan^2 \alpha > 2$.
As we discuss below, this condition is the same as
the condition for noise-induced turbulence in nonlocally coupled phase oscillators~\cite{ref:kawamura07}.

\section{On noise-induced turbulence} \label{sec:turbulence}

Finally, we briefly discuss the relation between ``effective anti-phase coupling'' and ``noise-induced turbulence''.
In this section, our arguments do not assume that collective oscillations are near the onset.
In Ref.~\cite{ref:kawamura07},
we considered a system of nonlocally coupled noisy phase oscillators described by the following model:
\begin{equation}
  \frac{\partial}{\partial t} \phi\left( \bd{r}, t \right)
  = \omega + \int d\bd{r}'\, G\left( \bd{r} - \bd{r}' \right)
  \Gamma\left( \phi\left( \bd{r}, t \right) - \phi\left( \bd{r}', t \right) \right)
  + \sqrt{D} \, \xi\left( \bd{r}, t \right),
  \label{eq:nonlocal-model}
\end{equation}
where $\phi({\bd r}, t)$ represents the phase field of spatially extended oscillatory media,
$G({\bd r})$ is a nonlocal kernel function that decays with the distance $|{\bd r}|$,
$\Gamma(\phi)$ is the phase coupling function,
$\xi({\bd r}, t)$ represents spatiotemporally white Gaussian noise, and $D$ is the noise intensity.

The Langevin phase equation~(\ref{eq:nonlocal-model})
can be transformed into a nonlinear Fokker-Planck equation in the following form:
\begin{align}
  \frac{\partial}{\partial t} f\left( \phi, \bd{r}, t \right)
  = &- \frac{\partial}{\partial \phi} \left[ \left\{
    \omega + \int_{0}^{2\pi} d\phi'\, \Gamma\left( \phi - \phi' \right)
    f\left( \phi', \bd{r}, t \right) \right\}
    f\left( \phi, \bd{r}, t \right) \right]
  + D \frac{\partial^2}{\partial \phi^2} f\left( \phi, \bd{r}, t \right) \nonumber \\
  &- \frac{\partial}{\partial \phi} \left[ 
    \int_{0}^{2\pi} d\phi'\, \Gamma\left( \phi - \phi' \right)
    \Bigl\{ G_{2} \nabla^2 f\left( \phi', \bd{r}, t \right) \Bigr\}
    f\left( \phi, \bd{r}, t \right) \right] \nonumber \\
  &- \cdots.
  \label{eq:nonlocal-nfp}
\end{align}
Here, we have expanded the nonlocal coupling term as
\begin{equation}
  \int d\bd{r}'\, G\left( \bd{r} - \bd{r}' \right) f\left( \phi', \bd{r}', t \right)
  = \sum_{n=0}^{\infty} G_{2n} \nabla^{2n} f\left( \phi', \bd{r}, t \right),
\end{equation}
where $G_{2n}$ is the $2n$-th moment of $G(\bd{r})$.

The space-dependent order parameter $A(\bd{r},t)$ is defined by
\begin{equation}
  A\left( {\bd r}, t \right)
  = R\left( {\bd r}, t \right) e^{i \Theta({\bf r}, t)}
  = \int d{\bd r}'\, G\left( {\bd r} - {\bd r}' \right)
  \int_0^{2\pi} d\phi'\, e^{i \phi'} f\left( \phi', {\bd r}', t \right),
\end{equation}
where $\Theta(\bd{r},t)$ can be considered as the space-dependent collective phase,
$f(\phi,\bd{r},t) = f_0(\phi - \Theta(\bd{r},t))$.
Applying the phase reduction method to Eq.~(\ref{eq:nonlocal-nfp}),
we obtained the following collective phase equation
\begin{equation}
  \frac{\partial}{\partial t} \Theta\left( \bd{r}, t \right)
  = \Omega + \bar{\nu} \nabla^2 \Theta\left( \bd{r}, t \right)
  + \bar{\mu} \bigl( \nabla \Theta\left( \bd{r}, t \right) \bigr)^2
  + \cdots,
  \label{eq:nonlocal-collective}
\end{equation}
where $\bar{\nu}$ and $\bar{\mu}$ are coefficients.
In particular, the collective phase diffusion coefficient $\bar{\nu}$ was given by
\begin{equation}
  \bar{\nu} = -G_2 \int_{0}^{2\pi} d\varphi \int_{0}^{2\pi} d\varphi' \,
  \Gamma\left( \varphi - \varphi' \right)
  k_0\left( \varphi \right) u_0\left( \varphi' \right),
  \label{eq:nu}
\end{equation}
which can be negative and then induce spatiotemporal chaos (turbulence).
Details of the definitions and the derivations are given in Ref.~\cite{ref:kawamura07}.

\begin{figure*}
  \begin{center}
    \includegraphics[width=0.50\hsize,clip]{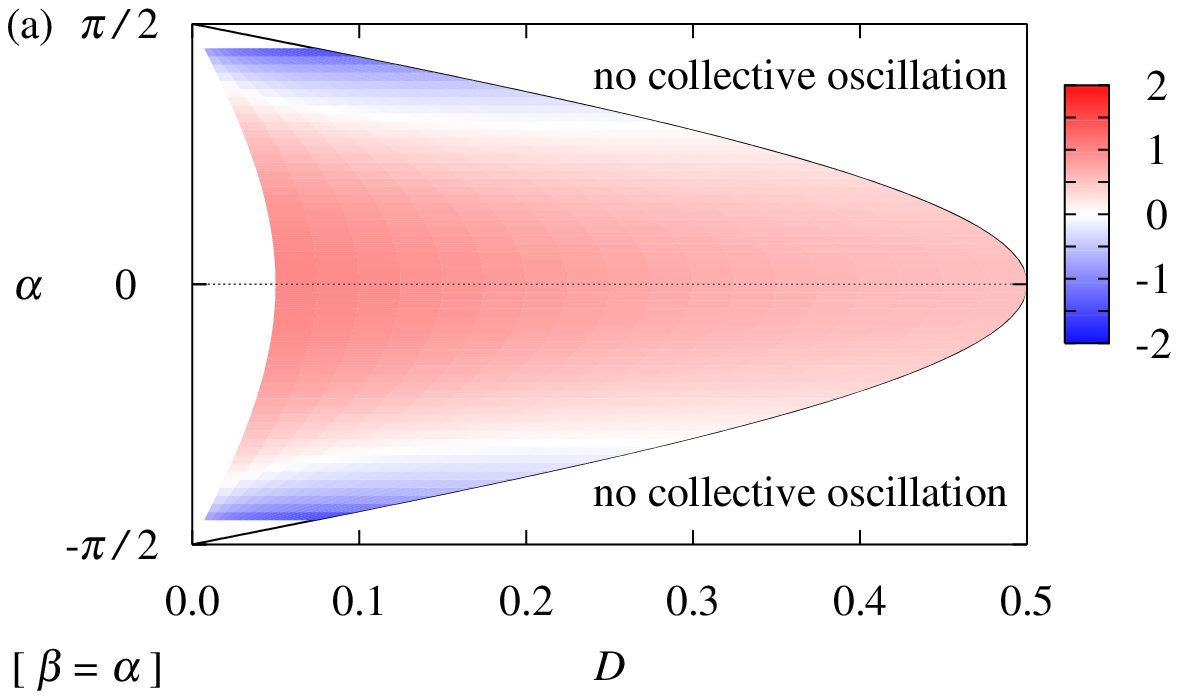}
    \includegraphics[width=0.42\hsize,clip]{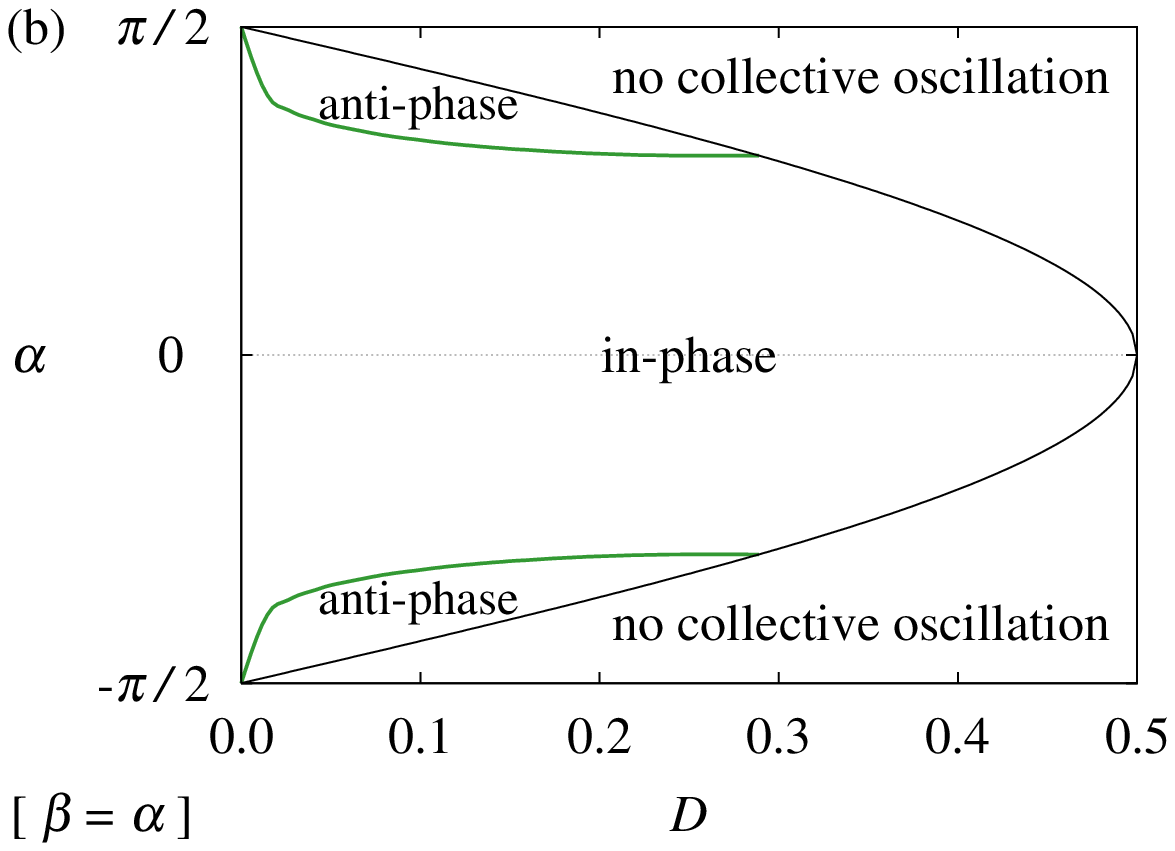}
    \caption{(Color online)
      Effective type of phase coupling between collective oscillations
      in $D$ and $\alpha$ with $\beta = \alpha$.
      (a) Dependence of $\rho \cos \delta$ on $D$ and $\alpha$.
      (b) The solid curves are determined by $\rho \cos \delta = 0$.
    }
    \label{fig:6}
  \end{center}
\end{figure*}

Now, it is clear that Eqs.~(\ref{eq:nfp}), (\ref{eq:collective}), and (\ref{eq:type})
describing interacting groups of globally coupled noisy phase oscillators are similar to
Eqs.~(\ref{eq:nonlocal-nfp}), (\ref{eq:nonlocal-collective}), and (\ref{eq:nu})
describing a system of nonlocally coupled noisy phase oscillators.
When the external phase coupling function is the same as the internal one,
i.e., $\Gamma_{\sigma\tau}(\varphi) = \Gamma(\varphi)$,
Eq.~(\ref{eq:type}) is essentially equivalent to Eq.~(\ref{eq:nu}).
Namely, the instability condition, $-\digamma'(0) < 0$, for in-phase collective synchronization between two groups,
which gives the anti-phase condition for the sinusoidal coupling,
coincides with the instability condition, $\bar{\nu} < 0$, for spatially uniform solutions of the collective phase equation.
Therefore, the phase diagram plotted in Fig.~\ref{fig:6} using $D$ and $\alpha$ with $\beta = \alpha$
is the same as that we obtained for noise-induced turbulence in Ref.~\cite{ref:kawamura07}.

The above situation for collective oscillations at the macroscopic level
is in parallel with the classical problem for phase oscillators at the microscopic level,
in which the instability condition for in-phase synchronization of two coupled phase oscillators
coincides with the instability condition for spatially uniform solutions of the phase diffusion equation~\cite{ref:kuramoto84}.

\section{Concluding remarks} \label{sec:remarks}

In the present paper,
we considered two weakly interacting groups of globally coupled noisy identical phase oscillators undergoing collective oscillations.
To analyze them, we adopted the idea of collective phase description,
namely, we treated the collective oscillations of each group as a single macroscopic phase oscillator.
We developed a theory that derives the collective phase coupling function between the groups
from the microscopic external phase coupling function between individual oscillator pairs belonging to the different groups.
Based on this theory, we illustrated counter-intuitive situations
in which the two groups become anti-phase synchronized despite in-phase microscopic coupling, and vice versa.
We also developed a theory that gives explicit analytical expressions of the collective phase coupling functions
near the onset of collective oscillations.
A complete phase diagram in the case of the sinusoidal internal and external coupling functions is summarized in Appendix.

In our companion work~\cite{ref:kawamura10},
we considered two weakly interacting groups of globally coupled noiseless non-identical phase oscillators
and discussed their collective synchronization properties.
The strong similarity in results between the two types of systems,
one stochastic and the other deterministically random, is remarkable,
while the theoretical methods employed are completely different between them.
In particular, we found the same counter-intuitive phenomena, namely,
the disagreement of the types between the collective phase coupling function
and the microscopic external phase coupling function.

The notion of collective phase description is convenient and powerful
in analyzing complex macroscopic rhythms arising from systems of interacting microscopic dynamical elements.
Further development of the theories will provide useful viewpoints
to understand various complex rhythms in real-world systems, in particular, their functional meaning.

\appendix

\section*{Appendix: phase diagram for sinusoidal coupling} \label{sec:appendix}

\begin{figure*}
  \begin{center}
    \begin{tabular}{c c}
      \includegraphics[width=0.50\hsize,clip]{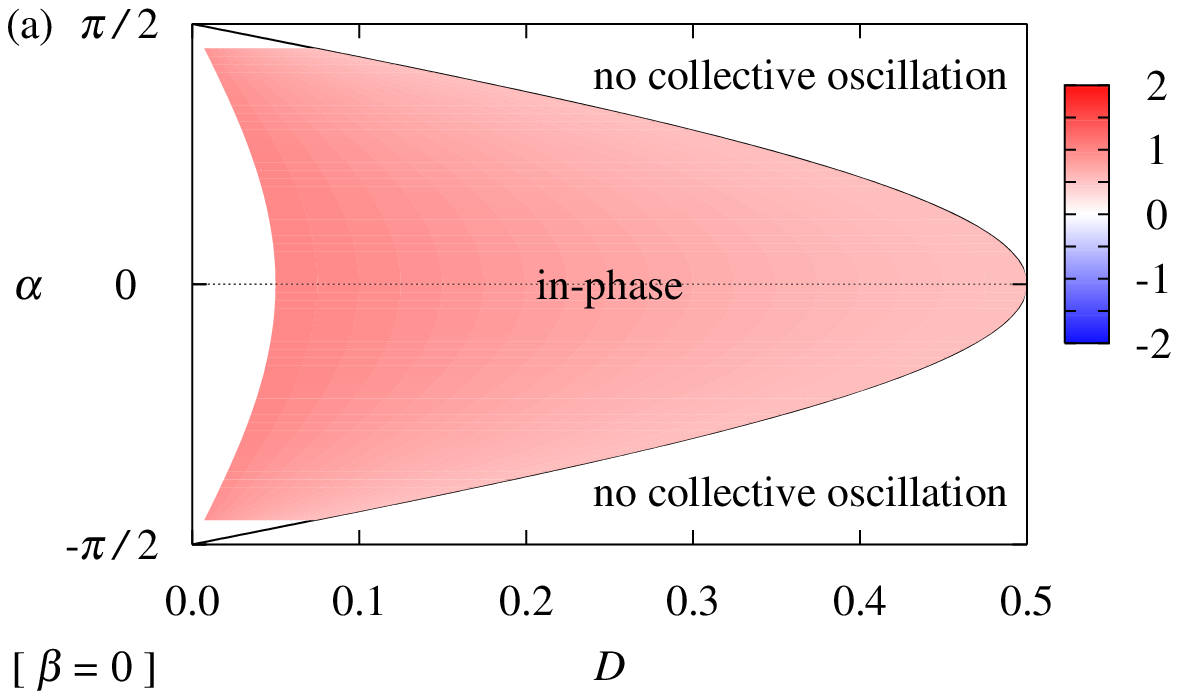} &
      \includegraphics[width=0.50\hsize,clip]{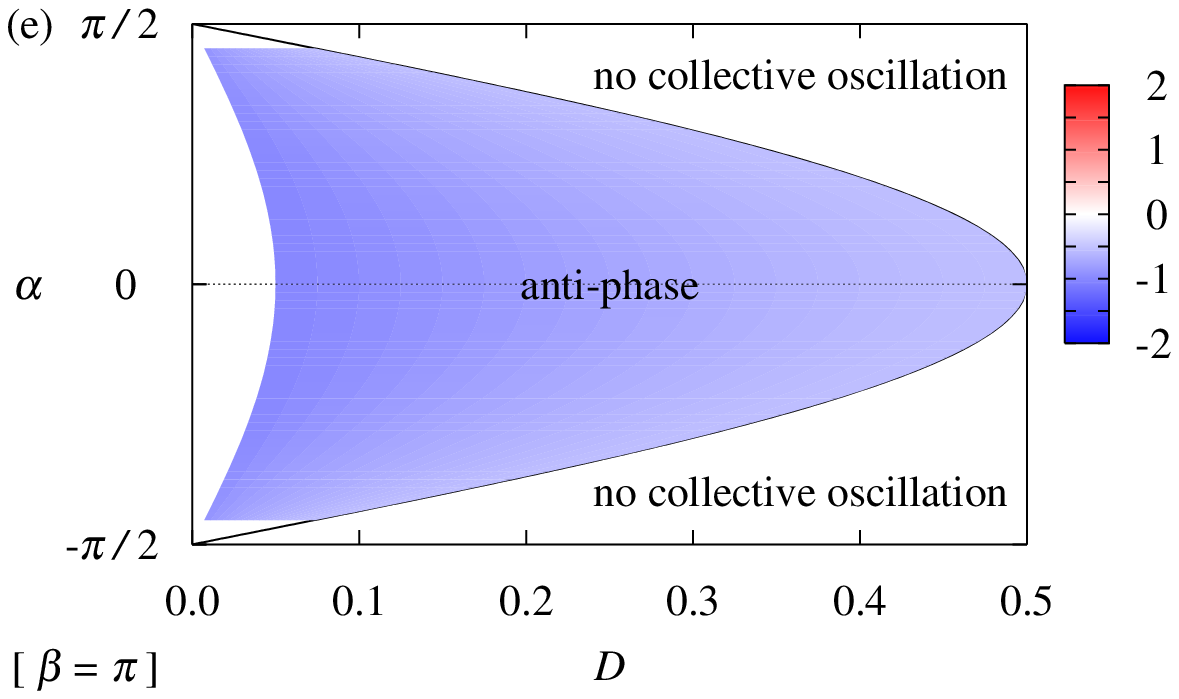} \\
      \includegraphics[width=0.50\hsize,clip]{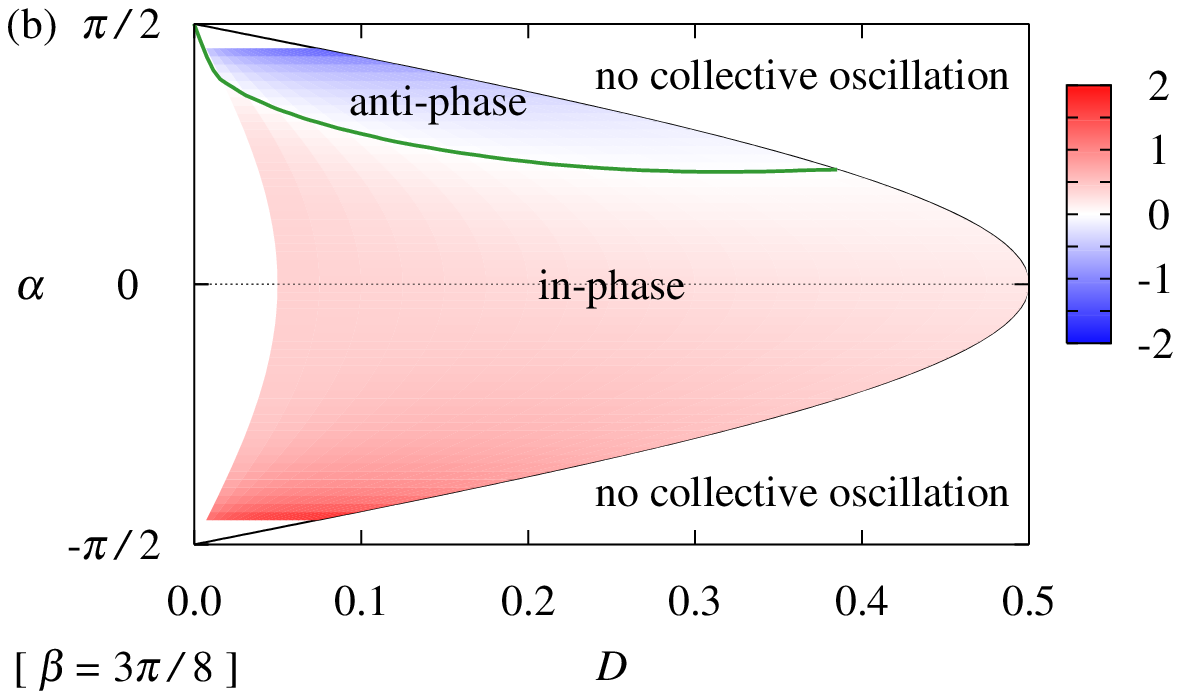} &
      \includegraphics[width=0.50\hsize,clip]{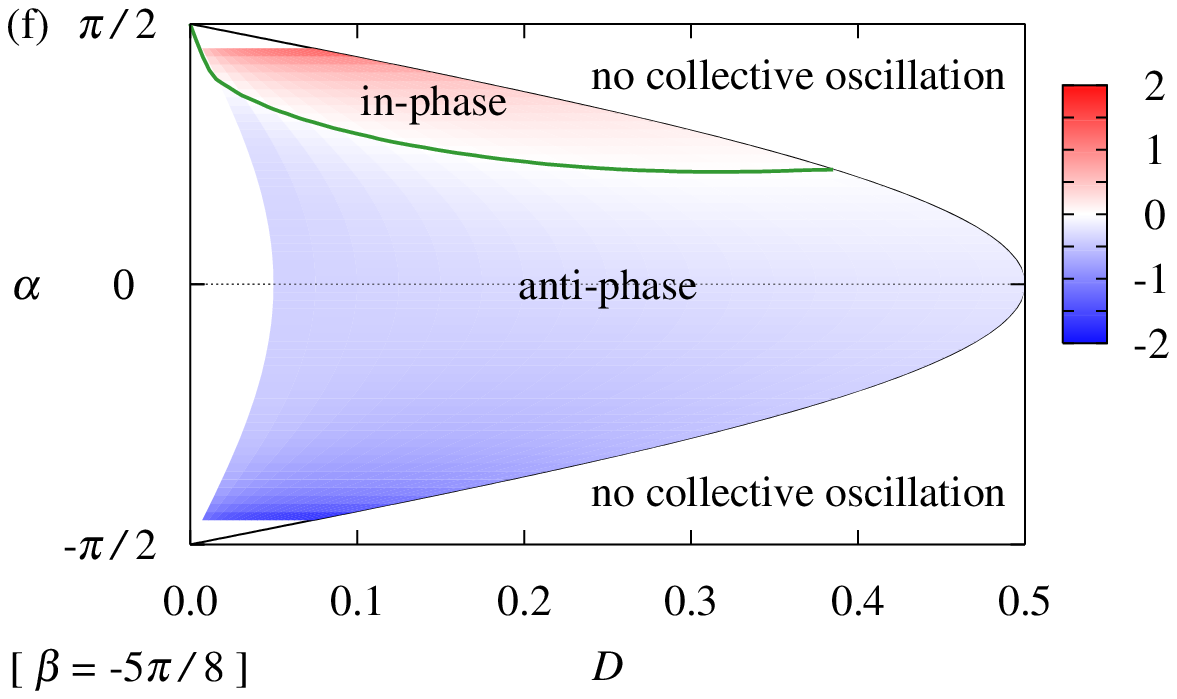} \\
      \includegraphics[width=0.50\hsize,clip]{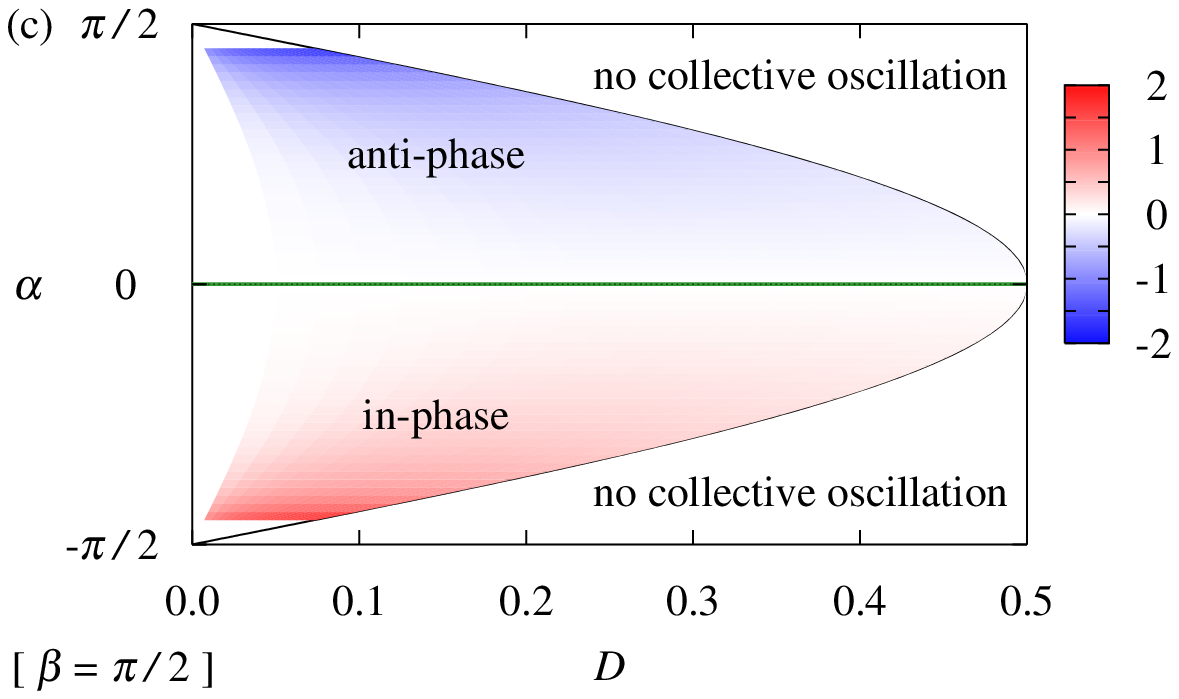} &
      \includegraphics[width=0.50\hsize,clip]{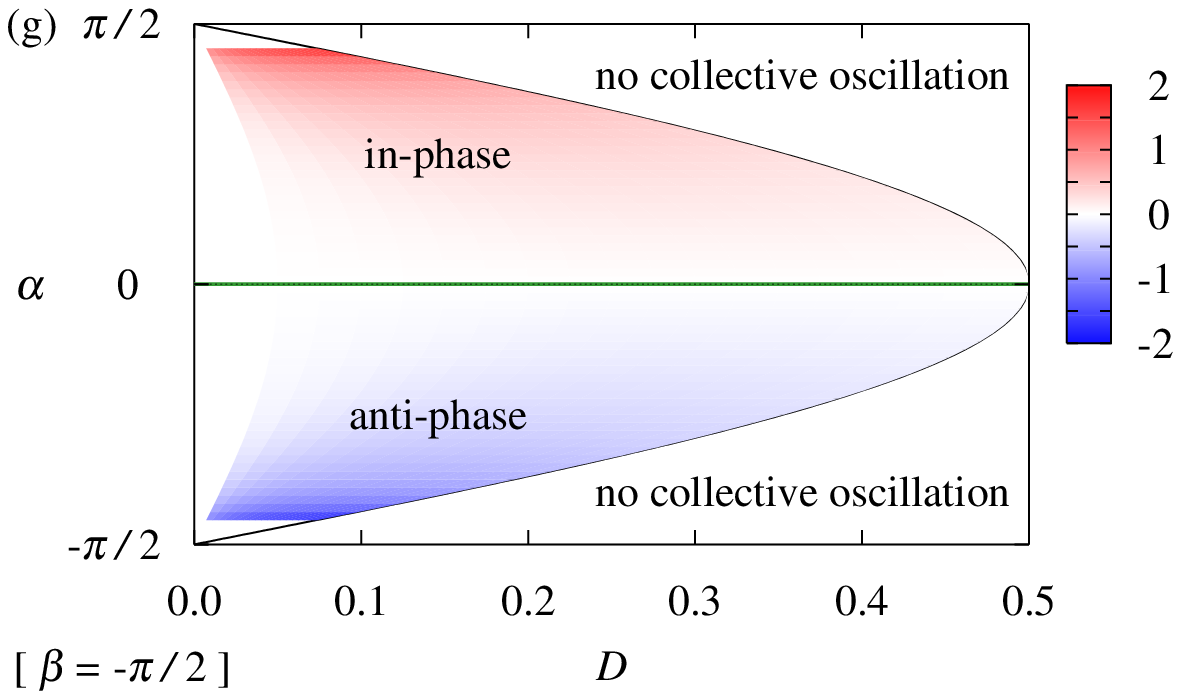} \\
      \includegraphics[width=0.50\hsize,clip]{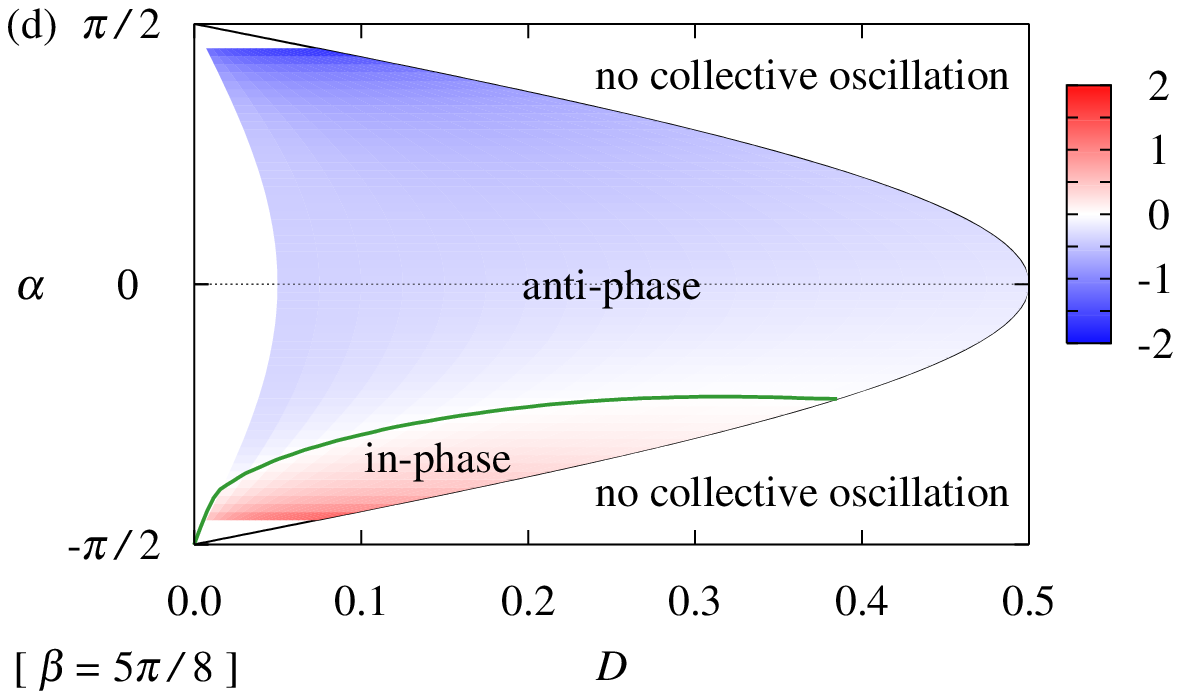} &
      \includegraphics[width=0.50\hsize,clip]{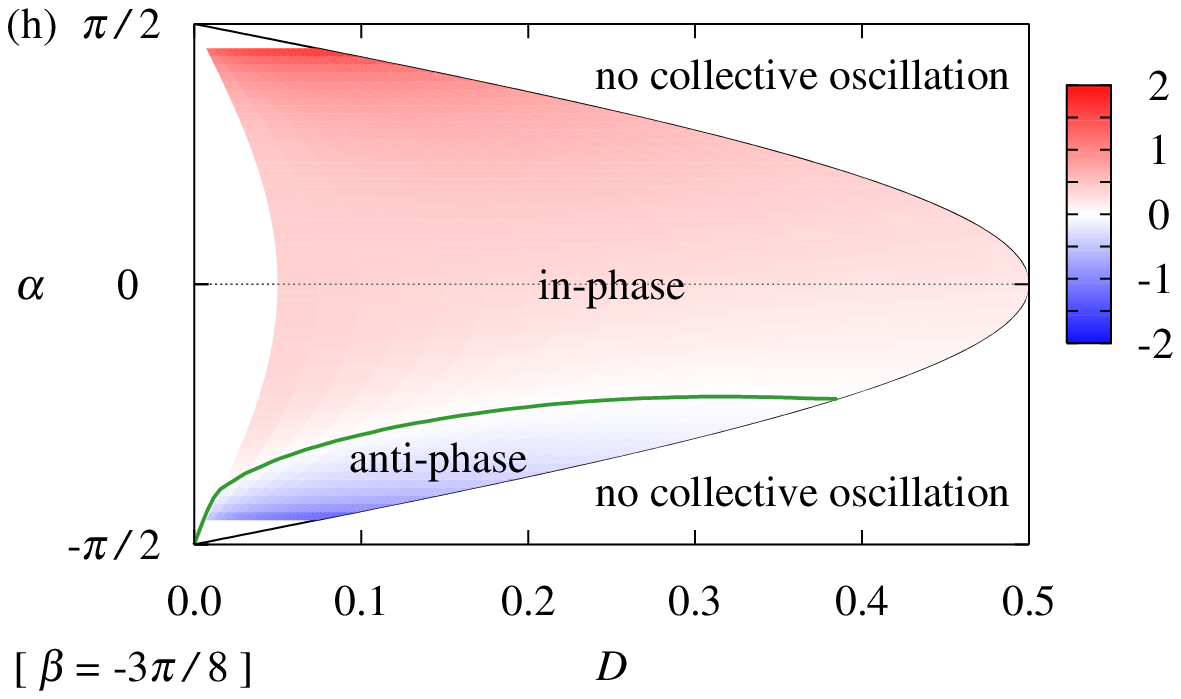} \\
    \end{tabular}
    \caption{(Color online)
      Phase diagram in $D$ and $\alpha$ with
      (a) $\beta = 0$,
      (b) $\beta = 3\pi / 8$,
      (c) $\beta = \pi / 2$,
      (d) $\beta = 5\pi / 8$,
      (e) $\beta = \pi$,
      (f) $\beta = -5\pi / 8$,
      (g) $\beta = -\pi / 2$, and
      (h) $\beta = -3\pi /8$.
    }
    \label{fig:A1}
  \end{center}
\end{figure*}

We here present a complete phase diagram for the case
of the sinusoidal internal and external coupling functions.
The type of the collective phase coupling function is found from 
$\rho \cos \delta$ given in Eq.~(\ref{eq:type}),
which was numerically evaluated for $\beta \in [-\pi, \pi]$
in the parameter region of $D / D_{\rm c} \in [0.1, 1.0)$ and
$\alpha \in [-a\pi/2, a\pi/2]$ with $a = 0.9$.
In addition, we used the analytical formula equation~(\ref{eq:type2})
on the Hopf bifurcation line, i.e., $D = D_{\rm c} = (\cos \alpha) / 2$.
Phase diagrams in $D$ and $\alpha$ with several values of $\beta$
are displayed in Fig.~\ref{fig:A1}.



\clearpage


\clearpage

\clearpage

\clearpage

\end{document}